\documentclass[traditabstract,a4]{aa}  

\usepackage{epsfig}
\usepackage{graphicx}
\usepackage{epstopdf}
\usepackage{color}  
\usepackage{array}   
\usepackage{units}  
\usepackage[breaklinks, colorlinks, citecolor=blue]{hyperref}
\usepackage{natbib}  
\usepackage{multirow}   
\usepackage{amsmath,amssymb}  
\usepackage[english]{babel}
\usepackage[latin1]{inputenc}
\usepackage[T1]{fontenc}
\usepackage{longtable,lscape}
\usepackage{footnote}

\usepackage{txfonts}
\usepackage[toc,page]{appendix} 

\begin{document}


\title{Cosmological constraints from the observed angular cross-power spectrum between Sunyaev-Zel'dovich and X-ray surveys}
\author{G.Hurier}
   \author{G. Hurier\inst{1}
   \and M. Douspis\inst{1}
      \and N. Aghanim\inst{1}
      \and E. Pointecouteau\inst{2,5}
      \and J.M. Diego\inst{3}
      \and J.F. Macias-Perez\inst{4} 
          }

\institute{Institut d'Astrophysique Spatiale, CNRS (UMR8617) Universit\'{e} Paris-Sud 11, B\^{a}timent 121, Orsay, France
\and CNRS; IRAP; 9 Av. colonel Roche, BP 44346, F-31028 Toulouse cedex 4, France
\and Instituto de F\`isica de Cantabria (CSIC-Universidad de Cantabria), Avda. de los Castros s/n, Santander, Spain
\and Laboratoire de Physique Subatomique et de Cosmologie, CNRS/IN2P3, Universit\'{e} Joseph Fourier Grenoble I, Institut National Polytechnique de Grenoble, 53 rue des Martyrs, 38026 
\and Université de Toulouse; UPS-OMP; IRAP; Toulouse, France\\
\\
\email{ghurier@ias.u-psud.fr} 
}

   \date{Received /Accepted}
 
   \abstract{
   
   We present the first detection of the cross-correlation angular power spectrum between the thermal Sunyaev-Zel'dovich (tSZ) effect and the X-ray emission over the full sky. 
   The tSZ effect and X-rays are produced by the same hot gas within groups and clusters of galaxies, which creates a naturally strong correlation between them that can be used 
   to boost the joint signal and derive cosmological parameters. 
   We computed the correlation  between the ROSAT All Sky Survey in the 0.5-2 keV energy band and the tSZ effect reconstructed from six Planck all-sky frequency maps between 70 and 545 GHz. 
   We detect a significant correlation over a wide range of angular scales. 
   In the range $50<\ell < 2000$, the cross-correlation
of X-rays to tSZ is detected at an overall significance of 28 $\sigma$. 
   As part of our systematic study, we performed a multi-frequency modelling of the AGN contamination and the correlation between cosmic infra-red 
   background and X-rays. 
   Taking advantage of the strong dependence of the cross-correlation signal on the amplitude of the power spectrum, we constrained $\sigma_8 = 0.804 \pm 0.037$, 
   where modelling uncertainties dominate statistical and systematic uncertainties. 
   We also derived constraints on the mass indices of scaling relations between the halo mass and X-ray luminosity, L$_{500}$-M$_{500}$, and SZ signal, Y$_{500}$-M$_{500}$, 
   $a_{\rm sz} + a_{\rm x} = 3.37 \pm 0.09$, and on the indices of the extra-redshift evolution, $\beta_{\rm sz} + \beta_{\rm x} = 0.4^{+0.4}_{-0.5}$.}


\authorrunning{Hurier et al.}
\titlerunning{Cosmological constraints from tSZ-X power spectrum}

   \maketitle

 
\section{Introduction}

Galaxy clusters are the largest virialized structures in the Universe. They are excellent tracers of the matter distribution, and their 
abundance can be used to constrain the cosmological model in an independent way.
Galaxy clusters can be identified in the optical bands as concentrations of galaxies \citep[see e.g.][]{abe89,gla05,koe07,ryk13}, they can be observed in X-rays by the bremsstrahlung emission 
produced by the ionized intra-cluster medium (ICM) \citep[see e.g.][]{boh00,ebe00,ebe01,boh01}. The same hot ICM also creates a distortion in the black-body spectrum of the cosmic 
microwave background (CMB) through the thermal Sunyaev-Zel'dovich (tSZ) effect \citep{sun69,sun72}, an inverse-Compton scattering between the CMB photons and the ionized electrons in the ICM. 
Recent catalogues based on measurements of the tSZ have been produced from {\it Planck} \citep{planckESZ,PlanckPSZ}, ACT \citep{mar11}, and SPT \citep{ble14} data.\\
The expected number of galaxy clusters is extremely sensitive to cosmological parameters, especially to the normalization of the matter power spectrum, 
or its fluctuations on 8 Mpc scales, $\sigma_8$.
Galaxy-cluster surveys have been used to constrain the cosmological parameters to a high level of precision  \citep{DES05,van10,seh11,boh14,PlanckSZC}.\\
Recent tSZ surveys, spanning wide areas in the sky, enable measuring the angular power spectrum of the tSZ signal over a wide range of multipoles and using it to constrain the cosmological model \citep{planckSZS}.  

The present constraints on cosmological parameters from tSZ power spectrum are limited by our knowledge of the relation between the total mass and the tSZ flux in galaxy 
clusters \citep{planckSL,has13,ben13}. 
An important factor that limits the accuracy in the determination of cosmological parameters is the contamination of the tSZ by other astrophysical emissions, mainly radio, infra-red point sources, 
and cosmic infra-red background \citep{dun11,shi11,rei12,sie13,planckSZS}.\\

Performing a power spectrum analysis in
the X-ray domain is difficult because the X-ray power spectrum is dominated by the signal from active galactic nuclei (AGN). 
Previous works have only been able to set upper limits on the cosmological model from the X-ray power spectrum alone \citep{die03}. \\
Since the X-ray emission and tSZ effect are produced by the same hot gas in galaxy clusters, we expect a strong correlation signal between tSZ and X-ray surveys.  
The correlation between tSZ and X-rays may reduce the contamination and systematic effects that affect each sample separately and set tighter limits on the  
cosmological model.\\
Recent studies \citep[see e.g.][]{hur14a} have shown that the higher sensitivity and resolution of the {\it Planck} mission
will probably allow us to detect a significant cross-correlation
of X-rays and tSZ.\\

The correlation between the tSZ signal and the X-ray emission has already been used in several studies. Statistical measurements of the tSZ effect have been obtained from WMAP 
data by stacking the temperature anisotropies toward known X-ray clusters \citep{afs05,lie06,afs07,atr08,die10,kom11,mel11,planckxsz} or by computing angular cross-power 
spectra between cluster catalogues and temperature anisotropy maps from WMAP \citep{fos03,ben03,mye04,her04,her06,hin07} and more recently from {\it Planck} \citep{haj13}.\\
Attempts have also been made to directly cross-correlate WMAP temperature maps with ROSAT All Sky Survey (RASS) \citep{die03}, but this was limited by the sensitivity and resolution 
of the WMAP experiment. 

The paper is organized as follows: 
In Sect.~\ref{modelxsz} we detail our modelling procedure for the cross-correlation of X-rays and tSZ. 
In Sect. \ref{secdata} we present the {\it Planck} and ROSAT data we used for this analysis. 
In Sect.~\ref{secmeth} we present our methodology for extracting the tSZ-X angular cross-correlation power spectrum
and show the first significant detection of the tSZ-X cross-correlation power spectrum from tSZ and X-ray full-sky data.
In Sect.~\ref{secerr} we estimate the uncertainty levels produced by data and modelling limitations. 
In Sect.~\ref{seccosmo} we show the constraints for cosmological and scaling law parameters.
We discuss our results in Sect.\ref{concl}.

\section{Modelling the tSZ-Xray cross-correlation}
\label{modelxsz}
\subsection{Thermal Sunyaev-Zel'dovich efffect}

The tSZ effect produces a small spectral distortion in the black-body spectrum of the CMB \citep{sun69,sun72}. Its intensity is related to the integral of the 
pressure along the line of sight,  or more specifically, the Compton parameter, which is defined as\begin{equation}
y = \int \frac{k_{\rm B} \sigma_T}{m_e c^2} n_e T_e {\rm d}l,
\end{equation}
where ${\rm d}l$ is the distance element along the line of sight
and $n_e$ and $T_e$ are the electron number density and the temperature.\\
In units of CMB temperature, the contribution of the tSZ effect to the sub-millimeter sky intensity for a given observation frequency $\nu$ is given by
\begin{equation}
\frac{\Delta T_{\rm CMB}}{T_{\rm CMB}} = g(\nu) y.
\end{equation}
Neglecting relativistic corrections, we have $g(\nu) = \left[ x {\rm coth} \left(\frac{x}{2}\right) - 4\right]$, with $x = h\nu/(k_{\rm B} T_{\rm CMB})$. This function is equal to 0 
around $217$~GHz, it is negative at lower frequencies and positive for higher frequencies. The characteristic signal of the spectral distortion from the ICM can be effectively used 
to directly isolate the projected pressure distribution in galaxy clusters from multi-frequency maps. 

\subsection{X-ray emission from galaxy clusters}

The ionized gas in the intra-cluster medium produces an X-ray emission via bremsstrahlung. This radiation is proportional to the square of the electron density. 
The energy spectrum of the X-ray emission from a galaxy cluster mainly depends on the temperature, $T_{500}$, of the ICM and to a lesser extent on the metallicity, $Z$, of the gas.
From an observational point of view, the X-ray spectrum also depends on the redshift, $z$.
The observed count rate at low energy depends on the column density of neutral hydrogen, $n_{\rm H}$, on the line of sight.
In this work, we model the galaxy clusters emission using a metal model \citep{mew85}.
We refer to \citet{hur14a} for a more detailed description of the X-ray emission modelling.

\subsection{Cross-correlation between tSZ effect and X-ray emission}

\label{secth}

The angular cross-power spectrum of tSZ effect and X-ray count map reads
\begin{equation}
C^{yX}_\ell = \frac{1}{2\ell +1} \sum_{m} \frac{1}{2}\left(y_{\ell m} x^{*}_{\ell m} + y^{*}_{\ell m} x_{\ell m}\right),
\end{equation}
with $y_{\ell m}$ and $x_{\ell m}$ the coefficients from the spherical harmonics decomposition of tSZ map and X-ray count map. This equation is applied without the detector transfer function and only for the full sky. Nevertheless, when computed for partial regions of the sky, we take the mask and beam effects into account, see Sect.~\ref{secmeth}.
To model this cross-correlation and the auto-correlation power spectra, we assume the following general expression
\begin{equation}
C^{yX}_{\ell} = C^{yX-{\rm 1h}}_\ell + C^{yX-{\rm 2h}}_\ell,
\end{equation}
where $C^{yX-{\rm 1h}}_\ell$ is the Poissonian contribution and $C^{yX-{\rm 2h}}_\ell$ is the two-halo term. 

The Poissonian term can be computed by assuming the square of the Fourier transform of normalized tSZ and X-ray projected profiles, weighted by the mass function and the respective tSZ effect flux and X-ray count-rate of galaxy clusters \citep[see e.g.][for a derivation of the tSZ auto-correlation angular power spectrum]{col88,kom02},
{\small
\begin{equation}
C_{\ell}^{yx-{\rm 1h}} = 4 \pi \int {\rm d}z \frac{{\rm d}V}{{\rm d}z {\rm d}\Omega}\int{\rm d}M \frac{{\rm d^2N}}{{\rm d}M {\rm d}V} (1+\rho_i \sigma_{{\rm log}\, Y}\sigma_{{\rm log}\, L})\overline{Y}_{500} \overline{S}_{500} y_{\ell} x_{\ell},
\end{equation}
}
where $\overline{S}_{500}$ and $\overline{Y}_{500}$ are the average X-ray count-rate and tSZ flux that depends on $M_{500}$ and $z$  . They are given by scaling relations \citep[see][]{hur14a}. ${{\rm d^2N}}/{{\rm d}M {\rm d}V}$ is the mass function of the
 dark matter halo (we considered here the fitting formula of \citet{tin08}), and ${{\rm d}V}/{{\rm d}z {\rm d}\Omega}$ is
the co-moving volume element. 
The factor $(1+\rho_i \sigma_{{\rm log}\, Y}\sigma_{{\rm log}\, L})$ accounts for the bias produced by the scatter of scaling relations \citep[see][]{hur14a}.\\
The Fourier transform of a 3D profile projected across the line
of sight on the sphere reads $\frac{4 \pi r_{\rm s}}{l^2_{\rm s}} \int_0^{\infty} {\rm d}x \, x^2 p(x) \frac{{\rm sin}(\ell x / \ell_{\rm s})}{\ell x / \ell_{\rm s}}$, where $p(x)$ is either tSZ or X-ray 3D profile, $x = r/r_{\rm s}$, $\ell_{{\rm s}} = D_{\rm ang}(z)/r_{\rm s}$, $r_{\rm s}$ is the scale radius of the profile.\\

The two-halo term corresponds to large-scale fluctuations of the dark matter field that induce correlations in the cluster distribution across the sky.
It can be computed as \citep[see e.g.][]{kom99,die04,tab11}
\begin{align}
C_{\ell}^{yx-{\rm 2h}} = 4 \pi \int {\rm d}z \frac{{\rm d}V}{{\rm d}z{\rm d}\Omega}&\left(\int{\rm d}M \frac{{\rm d^2N}}{{\rm d}M {\rm d}V} \overline{Y}_{500} y_{\ell} B(M,z)\right),\\ \nonumber
&\times \left(\int{\rm d}M \frac{{\rm d^2N}}{{\rm d}M {\rm d}V} \overline{S}_{500} x_{\ell} B(M,z)\right) P(k,z)
\end{align}
with $B(M,z)$ the time dependent linear bias that relates the matter power spectrum, $P(k,z)$, to the power spectrum of the cluster distribution across the sky. 
Following \citet{mo96} and \citet{kom99}, we adopt 
$$B(M,z)=1+(\nu^2(M,z)-1)/\delta_c(z),$$
with $\nu(M,z) = \delta_c(z)/\left[D_g(z) \sigma(M)\right]$, $D_g(z)$ is the linear growth factor and $\delta_c(z)$ is the over-density threshold for spherical collapse.\\
We stress that the two-halo term is negligible for tSZ-X-ray cross-correlation purposes.

The amplitude of the spectrum follows the dependencies \citep{hur14a} 
\begin{equation}
C_{\ell}^{yx-{\rm 1h}}  \propto \sigma_8^{8.12}  \Omega_{\rm m}^{3.42} H_0^{2.36} Y_\star L_\star (1-b)^{a_{\rm sz}+a_{\rm x}} \overline{N},
\end{equation}
where $Y_\star$, $L_\star$, are the normalizations of the $Y_{500}$-$M_{500}$ and $L_{500}$-$M_{500}$ relations, $a_{\rm sz}$, and $a_{\rm x}$ are the power-law indices ($Y_{500} \propto M^{a_{\rm sz}}_{500}$ and $L_{500} \propto M^{a_{\rm x}}_{500}$), $b$ is the bias between the mass estimated from X-ray measurements and the true matter mass of galaxy clusters. In the following we consider $b=0.20^{+0.10}_{-0.20}$ \citep[][and references therein]{PlanckSZC}. and $\overline{N}$ is the normalization of the mass function, for which we consider an uncertainty of 10\%.\\
We list the value and uncertainty on parameters of the $Y_{500}$-$M_{500}$ and $L_{500}$-$M_{500}$ scaling relations in Table.~\ref{tabscal}.
We define\begin{equation}
\Sigma_8 = \sigma_8  \left[\left( \frac{\Omega_{\rm m}}{0.32}\right)^{3.42} \left( \frac{H_0}{67}\right)^{2.36} \left(\frac{Y_\star}{0.65}\right) \left(\frac{L_\star}{1.88}\right) \left( \frac{1-b}{0.8}\right)^{3.43}  \left( \frac{\overline{N}}{\overline{N}_{0}}\right)\right]^{\frac{1}{8.12}}
\label{eqdep}
\end{equation}
as a single parameter that accounts for the overall amplitude
of the tSZ-X power spectrum.\\

\begin{table}
\center
\caption{Scaling-law parameters and error budget for the relations $Y_{500}-M_{500}$ \citep{PlanckSZC} and $L_{500}-M_{500}$ \citep{arn10}}
\begin{tabular}{|cc|cc|}
\hline
\multicolumn{2}{|c|}{$Y_{500}-M_{500}$} & \multicolumn{2}{c|}{$L_{500}-M_{500}$} \\
\hline
${\rm log}\,Y_\star$ & -0.19 $\pm$ 0.02 & ${\rm log}\, L_\star$ & 0.724 $\pm$ 0.032   \\
$\alpha_{\rm sz}$ & 1.79 $\pm$ 0.08 & $\alpha_{\rm x}$ & 1.64 $\pm$ 0.12   \\
$\sigma_{{\rm log}\, Y}$ & 0.075 $\pm$ 0.010 & $\sigma_{{\rm log}\, L}$ & 0.183 $\pm$ 0.032  \\
\hline
\end{tabular}
\label{tabscal}
\end{table}

\section{Data}
\label{secdata}

We used the {\it Planck} nominal mission dataset \citep{PlanckOVER} available at {\it Planck} Legacy Archive (PLA\footnote{http://www.sciops.esa.int}). 
We considered frequencies from 70 to 857 GHz. The two lowest frequency channels, at 30 and 44 GHz, were not considered because
their angular resolution is too poor.
We assumed that the {\it Planck} beams can be well approximated by circular Gaussian beams. We considered FWHM values from \citet{planckBEAMS}.
For the tSZ transmission in {\it Planck} spectral bandpasses, we used the values provided in \citet{planckBP}.\\

We used the ROSAT all-sky survey (RASS) public data\footnote{ftp://ftp.xray.mpe.mpg.de/rosat/archive/}, which cover 99.8\% of the sky, including 97\% that have an exposure time longer than 100s \citep{vog99}.
X-ray photons with an energy below 0.5 keV were not considered to reduce the impact of $n_{\rm H}$ absorption (considering the
photoelectric cross-section from \citet{mor83} and the typical $n_{\rm H}$ value from \citet{kal05}).
Then, we constructed a full-sky map of the photon count rate in the energy range 0.5-2.0 keV from each ROSAT photon event file and exposure map. 
We projected each event over the sky using {\tt HEALPix} \citep{gor05} pixelisation scheme at a resolution of $N_{\rm side} = 2048$.
The ROSAT exposure maps were reprojected using a nearest-neighbour interpolation on an {\tt HEALPix} grid with $N_{\rm side} = 2048$. 
Thus, the reprojected full-sky RASS has a resolution of 1.7 arcmin (size of the {\tt HEALPix} pixels for $N_{\rm side} = 2048$). Below, we account for the loss of power produced by the convolution of the RASS data with the pixel window function.\\
We have checked that reprojecting ROSAT number count images provides similar results for our analysis as would be achieved with ROSAT photon event files.\\

\begin{figure}[!th]
\begin{center}
\includegraphics[angle=90,scale=0.3]{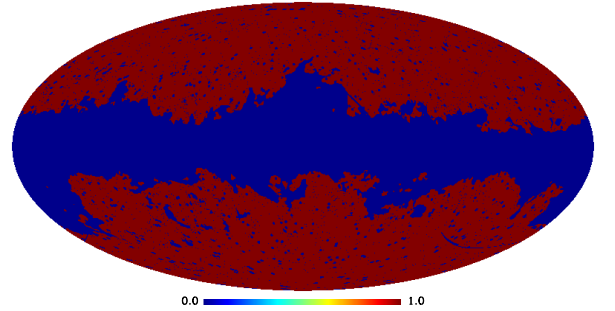}
\caption{60\% sky mask used to compute the tSZ-X cross-power spectrum.}
\label{mask}
\end{center}
\end{figure}

We also constructed a mask to avoid contamination by galactic or point source emissions in Planck data.
Considering that the {\it Planck} 857 GHz channel is a good tracer of the thermal dust emission, we chose to mask all regions that present an emission above 3 $K_{\rm CMB}$\footnote{$K_{\rm CMB}$ is defined as the unit in which a black body spectrum at 2.725~K is flat with respect to the frequency.} at this frequency \citep[see][for unit convention and conversion]{planckBP}. 
We masked all sources from the {\it Planck} compact source catalogue \citep{planckPCC} detected in at least one frequency with a signal-to-noise
ratio (S/N) above 5.
We also masked all regions with an exposure of 0 seconds in the RASS survey. After applying these cuts, we kept about 60\% of the sky for the analysis. We present the resulting mask in Fig.~\ref{mask}. We also considered masks of 20\%\ and 40\% for robustness checks with different thresholds for the thermal dust emission.\\

\begin{figure*}[!th]
\begin{center}
\includegraphics[scale=0.2]{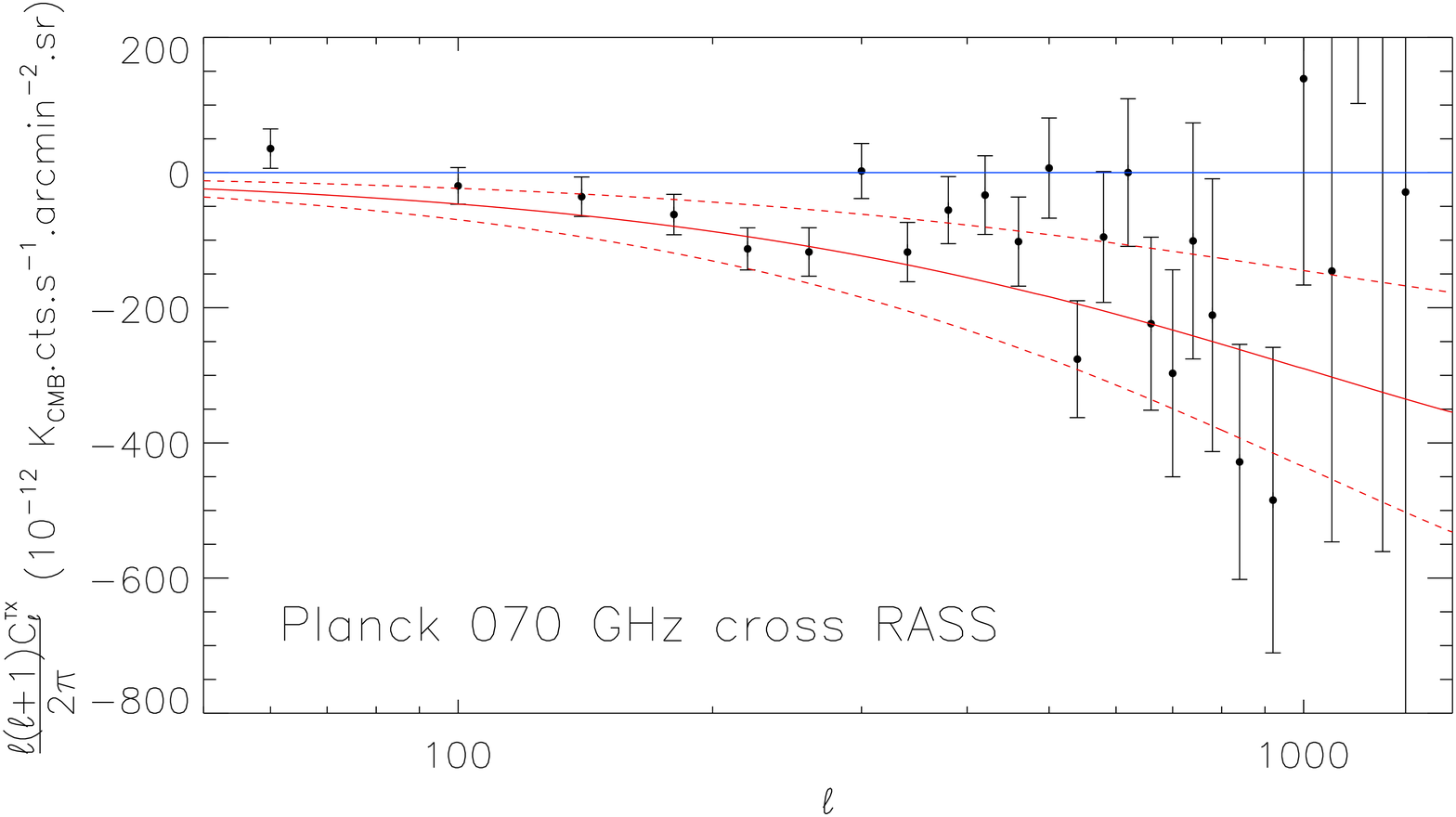}
\includegraphics[scale=0.2]{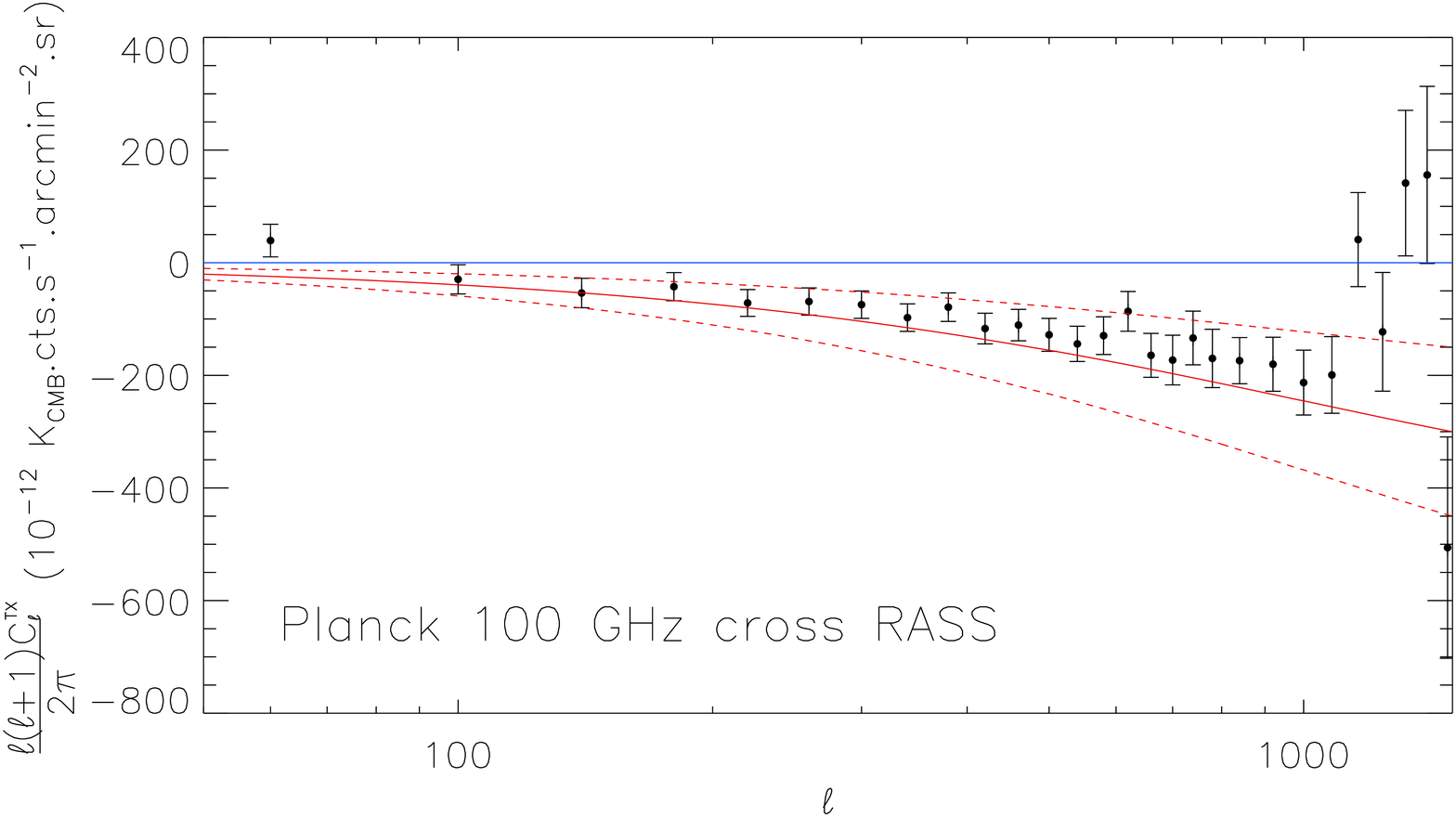}
\includegraphics[scale=0.2]{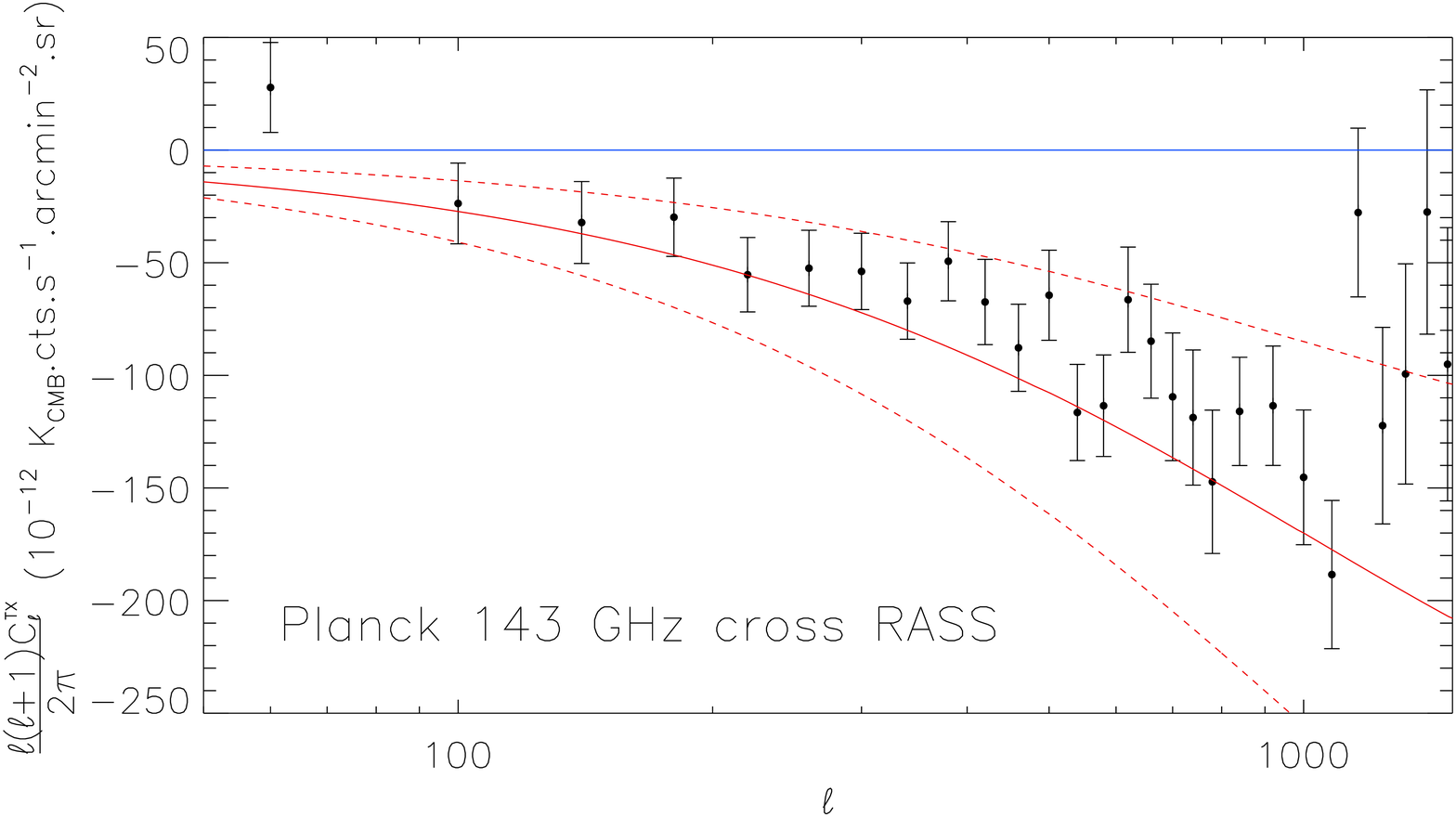}
\includegraphics[scale=0.2]{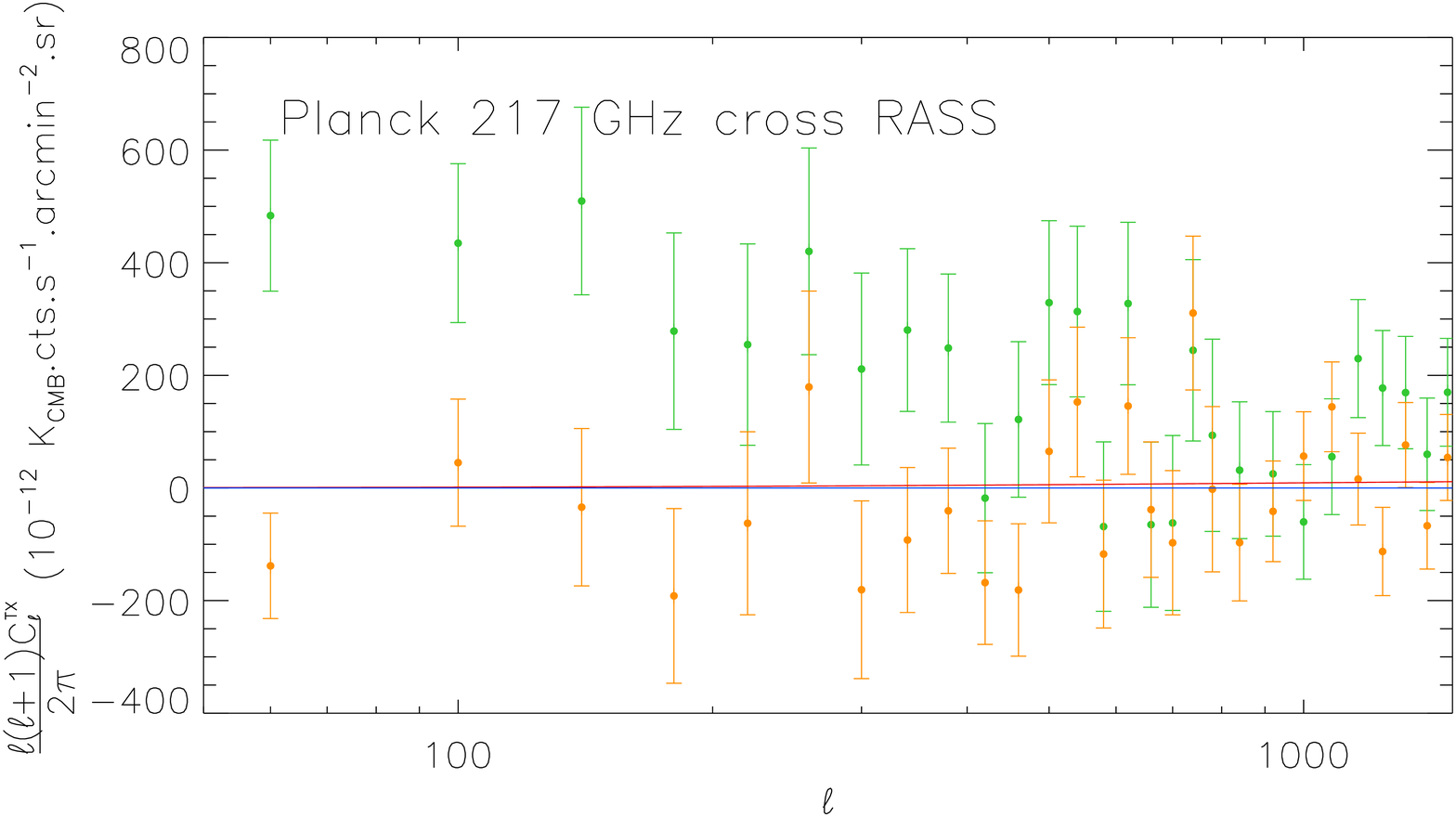}
\includegraphics[scale=0.2]{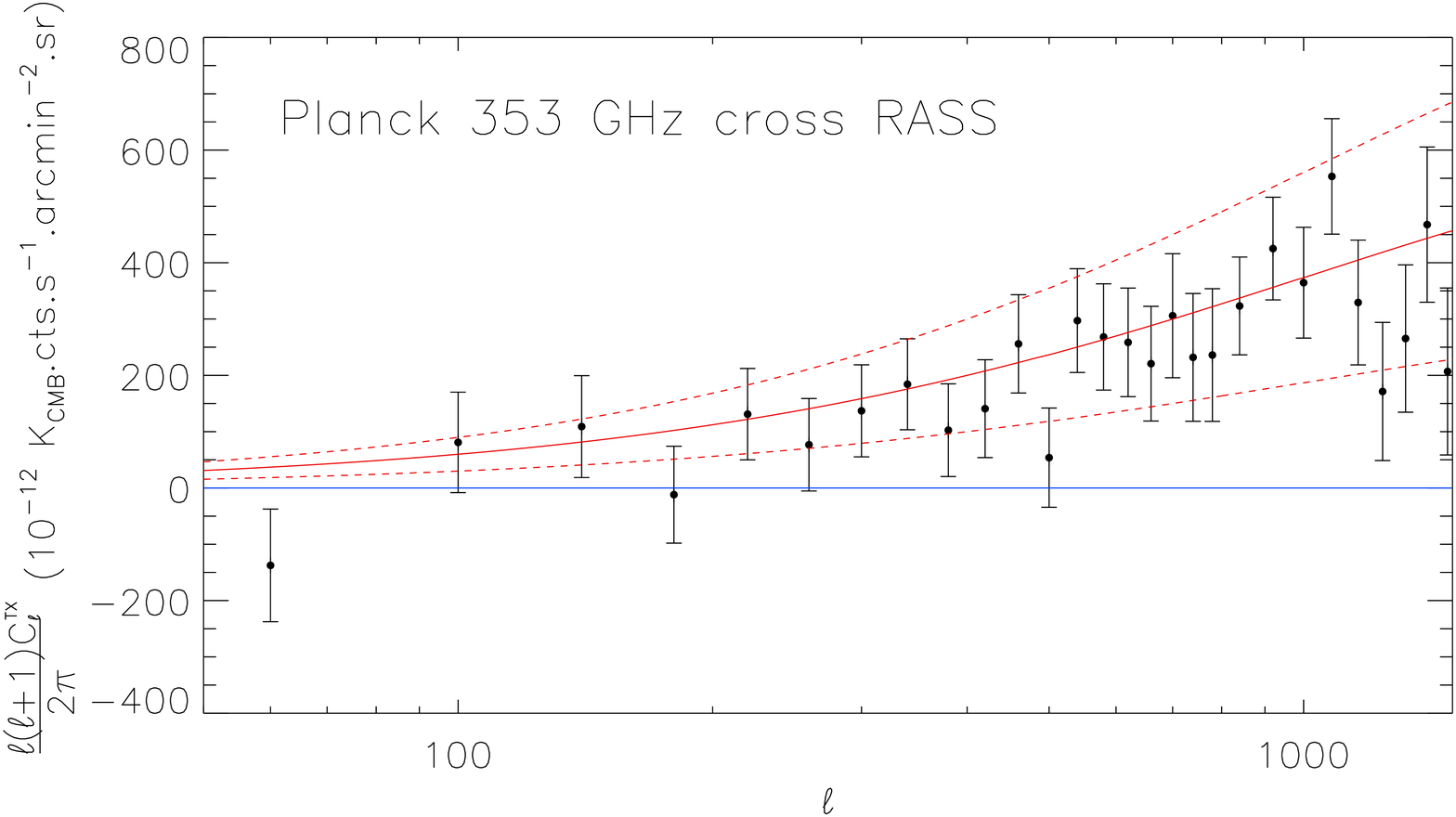}
\includegraphics[scale=0.2]{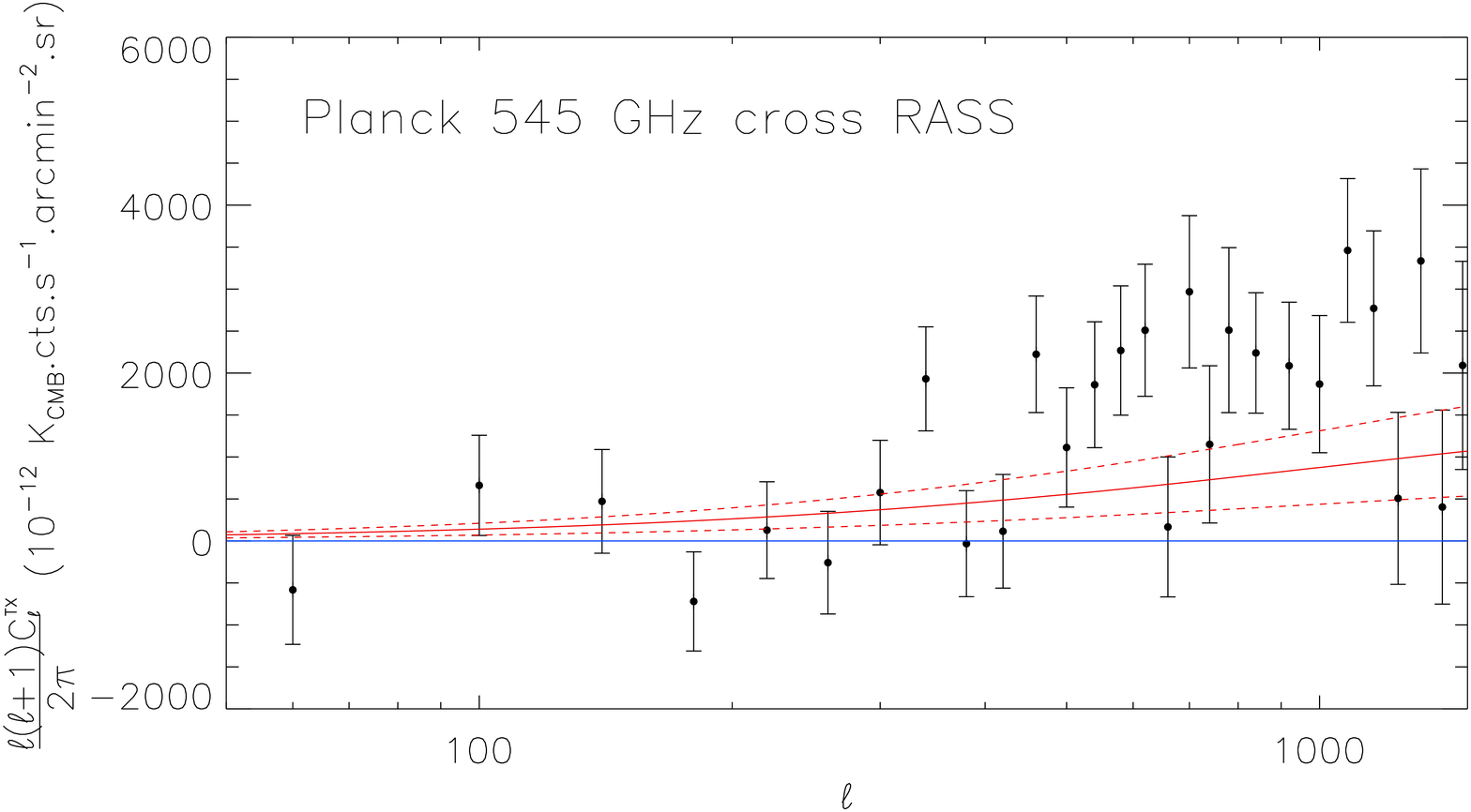}
\caption{Cleaned cross-correlation power spectra between the RASS in the 0.5-2.0 keV energy band and the {\it Planck} frequency maps. From left to right and top to bottom we display spectra
for 70, 100, 143, 217, 353 and 545 GHz. Data samples and error bars at 1 $\sigma$ are presented in black, the theoretical prediction using the {\it Planck} best-fitting cosmological parameters is plotted as a red solid line, the 1 $\sigma$ uncertainties on the theoretical prediction as estimated in \citet{hur14a}
are depicted as red dashed lines, the cross-correlation at 217 GHz before CMB and dust subtraction is displayed in green, and the cross-correlation at 217 GHz only corrected for dust in orange. Uncertainties have been computed as presented in Eq.~\ref{eqerrtx}}.
\label{crossfreqbestfit}
\end{center}
\end{figure*}

We chose to exclude the {\it Planck} official CMB and foreground maps from the cleaning of frequency maps. This choice was motivated by the contamination of the {\it Planck} official CMB maps by tSZ residuals \citep[see e.g.][]{bob14}, which would bias our analysis. We also show in Sect.~\ref{secperfreq} that galactic foregrounds can be kept under control using thermal dust cleaning and a galactic mask. As a consequence, we did not perform CO \citep{PlanckCO} or low-frequency galactic emission \citep{planckCOMP} cleaning, because it would have increased the noise level without significantly improving the results.

\section{Dectecting the tSZ-Xray cross-correlation}
\label{secmeth}

The tSZ-X angular cross-power spectrum can be estimated from frequency maps or from a tSZ Compton parameter map.
In this section, we explore the two approaches and discuss advantages and drawbacks of each of them.\\
In each case, we binned the power spectra and corrected the cross-spectra for beam and mask effects. The beam was corrected for by dividing the power-spectra by the beam transfer function in $\ell$ space. We deconvolved the power spectra by the mask-induced mixing matrix, which also accounts for the covered sky fraction.
For a detailed description of the mask correction see \citet{tri05}.

\subsection{Analysis for each frequency channel}
\label{secperfreq}

We computed the angular cross-power spectrum, $C_\ell^{\nu,{\rm RASS}}$, between {\it Planck} frequency maps from $\nu = $ 70 to 857 GHz and the RASS full-sky map in the 0.5-2.0 keV energy band. 

At a given frequency, the sky signal is dominated by CMB below 217 GHz and by thermal dust emission above 217 GHz.

We first corrected for CMB contamination by applying the transformation
\begin{equation}
\tilde{C}_\ell^{\nu,{\rm RASS}} = \frac{C_\ell^{\nu,{\rm RASS}} - C_\ell^{217,{\rm RASS}}}{1 - f(217)/f(\nu)}
,\end{equation}
where $f(\nu)$ is the tSZ transmission in the {\it Planck} map at the frequency, $\nu$. 
The term $f(217)/f(\nu)$ is very small because the tSZ transmission in the {\it Planck} 217 GHz channel is almost null.
This cleaning process prevents contamination at calibration uncertainty level \citep{PlanckCAL} by processes that follow the same spectral energy distribution (SED) as the CMB black-body radiation (e.g. the kinetic SZ effect).\\
Then we corrected for thermal dust contamination by applying the transformation
\begin{equation}
\widehat{C}_\ell^{\nu,{\rm RASS}} = \frac{\tilde{C}_\ell^{\nu,{\rm RASS}} - \left(\rho_{857,\nu} - \rho_{857,217} \right)C_\ell^{857,{\rm RASS}}}{1 - \left(\rho_{857,\nu} - \rho_{857,217} \right)f(857)/f(\nu)}.
\end{equation}
The value of $\rho_{857,\nu}$ was estimated from a linear fit of the lowest multipoles of the power spectra of ($\ell < 10$)
because the thermal dust emission dominates the cross-power spectra at low multipoles.
The factor $\left(\rho_{857,\nu} - \rho_{857,217} \right)f(857)/f(\nu)$ is negligible because of the ratio between tSZ and thermal dust SEDs at high frequencies.
$\widehat{C}_\ell^{\nu,{\rm RASS}}$ spectra are referred to as the cleaned cross-power spectra in the following.\\

\begin{figure}[!th]
\begin{center}
\includegraphics[scale=0.2,trim = 0cm 0cm 0cm 1.3cm, clip]{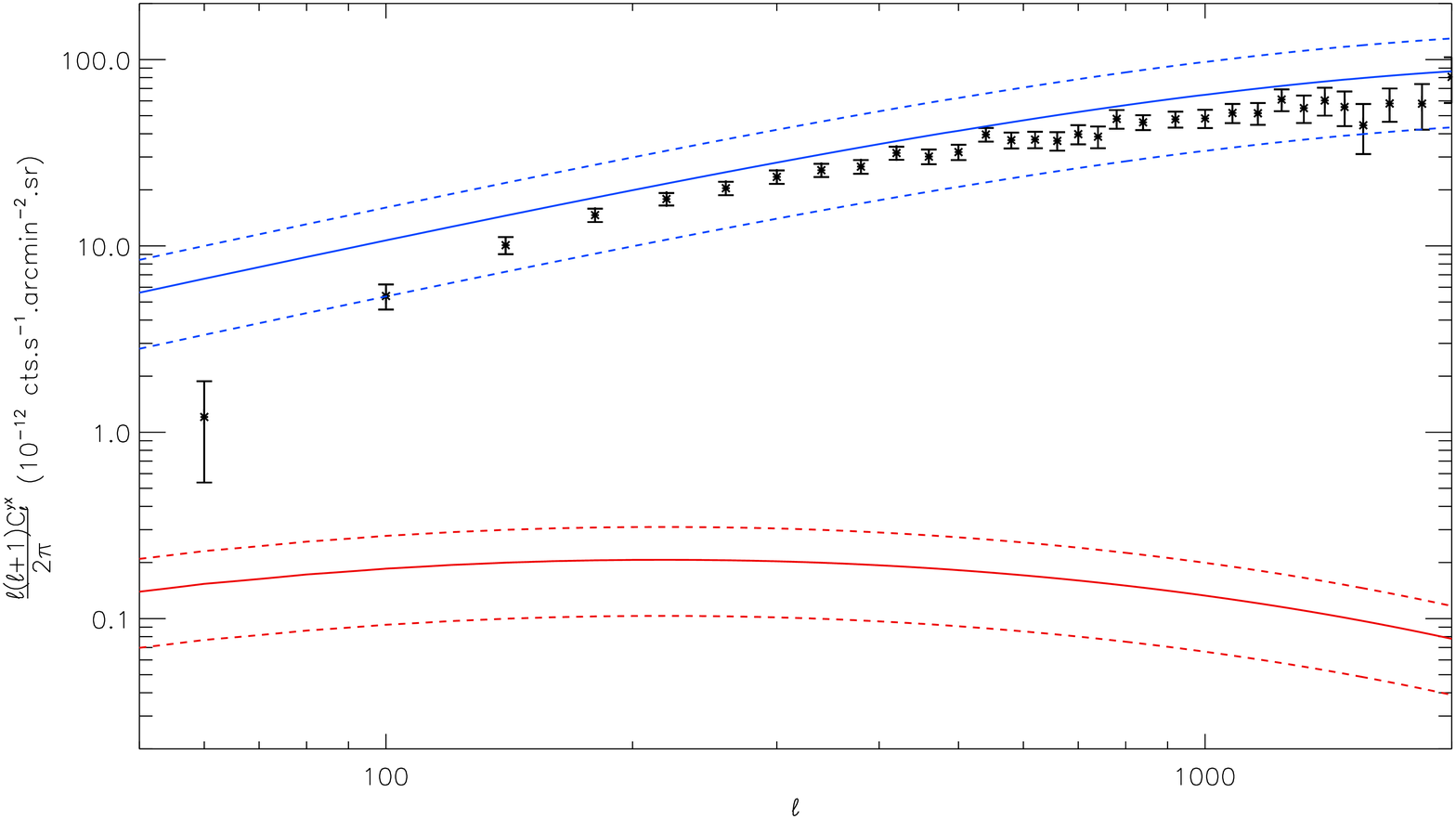}
\caption{In black we present the cross-correlation power spectrum between the RASS in the 0.5-2.0 keV energy band  map and the MILCA $y$-map computed from {\it Planck} data. As a dark blue solid line we show the theoretical one-halo tSZ-Xray cross-correlation power spectrum for {\it Planck} best-fitting cosmological parameters, in red the contribution of the two-halo term, and as dashed lines we show the 1 $\sigma$ uncertainties on the theoretical prediction as estimated in \citet{hur14a}. Uncertainties have been computed as presented in Eq.~\ref{eqerryx}.}.
\label{crossybestfit}
\end{center}
\end{figure}

In Fig.~\ref{crossfreqbestfit}, we present the cleaned cross-power spectra between the RASS map and the {\it Planck} maps at 70, 100, 143, 353, and 545 GHz.
We computed the $\chi^2$ of each spectrum compared to a null correlation. We obtain $\chi^2 = 86$, $274$, $378$, $205$, and $134$ for 31 degrees of freedom from $\ell = 100$ to $\ell = 2000$ at 70, 100, 143, 353, and 545 GHz, respectively. We also display the cross-correlation at 217 GHz without CMB and dust subtraction, and with dust correction alone. After dust correction, the cross-spectrum at 217 GHz is compatible with zero, with $\chi^2 = 30$ for 31 degree of freedom.\\
We compared these cross-spectra with the theoretical prediction assuming the {\it Planck}-CMB best-fitting $\Lambda$CDM cosmology ($\Omega_{\rm m} = 0.3175$, $\sigma_8 = 0.8344$, and $H_0 = 67$ km/s/Mpc) \citep{planckPAR}. 
We observe a significant correlation that follows a tSZ SED. 
However, we note that the correlation signal appears to be significantly weaker than the prediction.
We also note a significant contamination in the cross-power spectra at 100 and 143 GHz. This contamination can be seen at high multi-pole values ($\ell$ > 1000) and is interpreted as a correlation between radio and X-ray AGNs that compensates the negative correlation between the tSZ and the X-ray emission from galaxy clusters at those frequencies.\\
The angular cross-power spectrum at 545 GHz significantly exceeds the theoretical prediction. This excess can be interpreted as a contamination by the correlation between X-ray emission and the cosmic infra-red background.\\
Figure~\ref{crossfreqbestfit} illustrates the main advantages of using a multi-frequency measurement of the tSZ-X cross-power spectrum. Indeed, with multi-frequency measurements, we can easily distinguish contamination sources from the tSZ-X signal by using the differences in SEDs.

\subsection{Compton parameter map based analysis}
\label{secymap}

We used a second approach to extract the tSZ-X cross-power spectrum. 
In this case, we built a tSZ Compton parameter map ($y$-map) using the MILCA method \citep{hur13a} on the {\it Planck} maps from at 100 to 857 GHz. 
We verified that including frequencies from 30 to 70 GHz does not change the results.\\ 
The tSZ-X angular cross-power spectrum is obtained by directly cross-correlating the reprojected RASS full-sky map and the $y$-map, using 69\% of the sky.\\

In Fig.~\ref{crossybestfit}, we present the derived cross-power spectrum and compare it with the theoretical prediction assuming the Planck-CMB best-fitting cosmology. 
We also computed the $\chi^2$ of the spectrum with respect to a null correlation. We derive $\chi^2 = 2687$ for 33 degrees of freedom from $\ell = 50$ to $\ell = 2000$.
Similarly to the multi-frequency analysis, we observe that the measured power spectrum is significantly weaker than the prediction. 
However, the shape of the measured power spectrum agrees with the shape of the prediction. 
As the $y$-map reconstruction is performed both in pixel and frequency domains, it allows us to extract the tSZ signal at a higher S/N than a linear combination of power spectra.
However, the identification and estimate of contamination sources is more complicated.\\
Consequently, the two approaches are complementary and were used together to check the robustness of the results.

\subsection{Robustness of the detection}

In this section we verify that our signal is produced by galaxy
cluster and not by AGNs.

\begin{figure}[!ht]
\begin{center}
\includegraphics[scale=0.2]{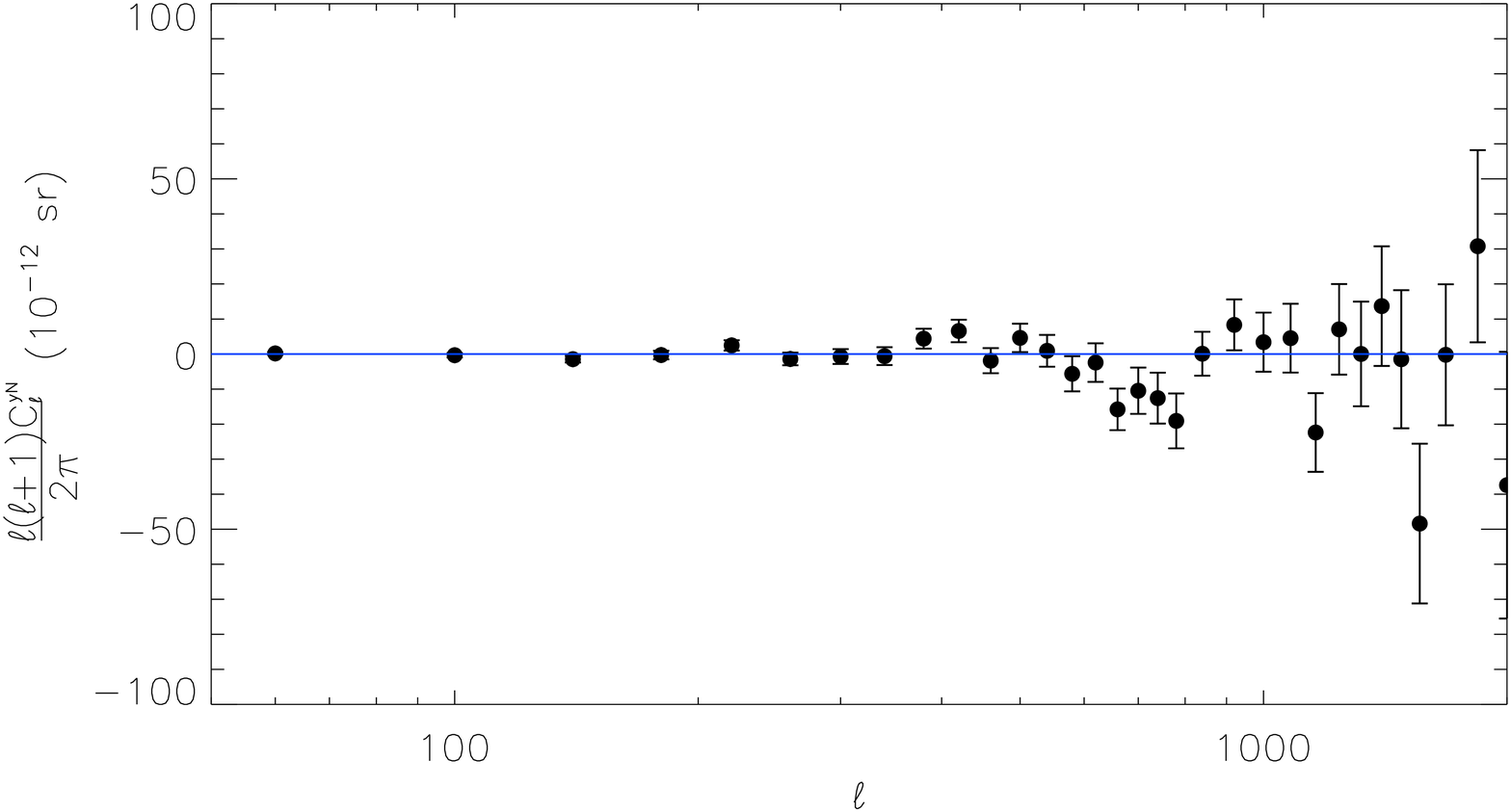}
\includegraphics[scale=0.2]{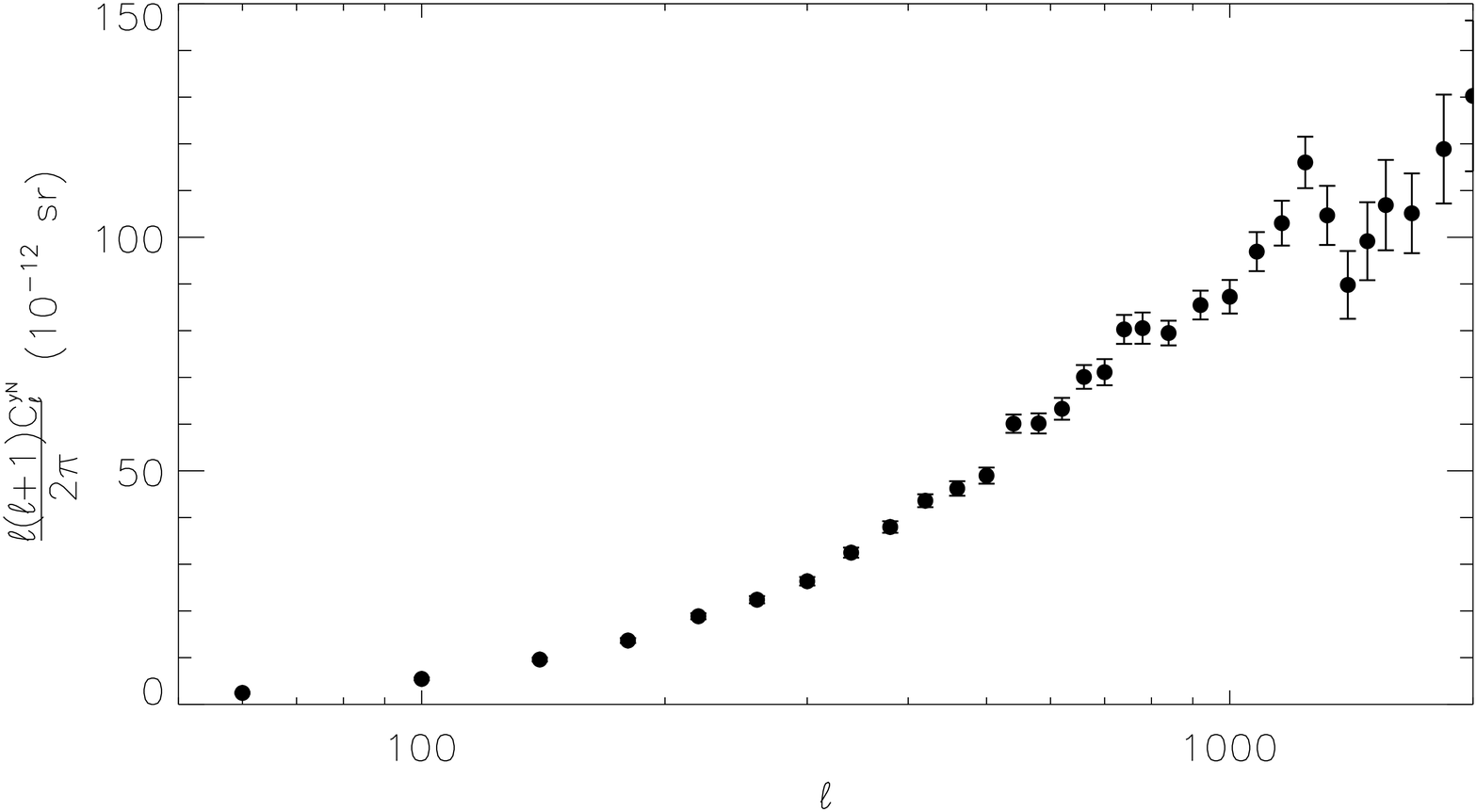}
\caption{Top panel: cross-spectrum between the MILCA $y$-map and the RBS sources flagged as AGN (black samples), the blue line shows the 0 correlation. Bottom panel: cross-spectrum between the MILCA $y$-map and the RBS sources flagged as galaxy clusters.}
\label{check}
\end{center}
\end{figure}

We estimated the contamination from known AGN using X-ray catalogues such as the ROSAT Bright Survey \citep[RBS,][]{fis98}.
First, we projected the RBS sources on a full-sky map.
We computed the cross-spectrum between the $y$-map and the projected RBS map considering only sources flagged as AGNs. Then, we performed the same analysis for sources flagged as clusters.
Uncertainties were computed following the approach presented in Sect.~\ref{secerr}.
We present the obtained cross-spectra in Fig.~\ref{check}. We do not observe any significant contribution from this AGNs sample. These spectra illustrate that our signal is dominated by X-ray emission from galaxy clusters and not from X-ray AGNs.\\

\begin{figure}[!th]
\begin{center}
\includegraphics[scale=0.2,trim = 0cm 0cm 0cm 1.3cm,clip]{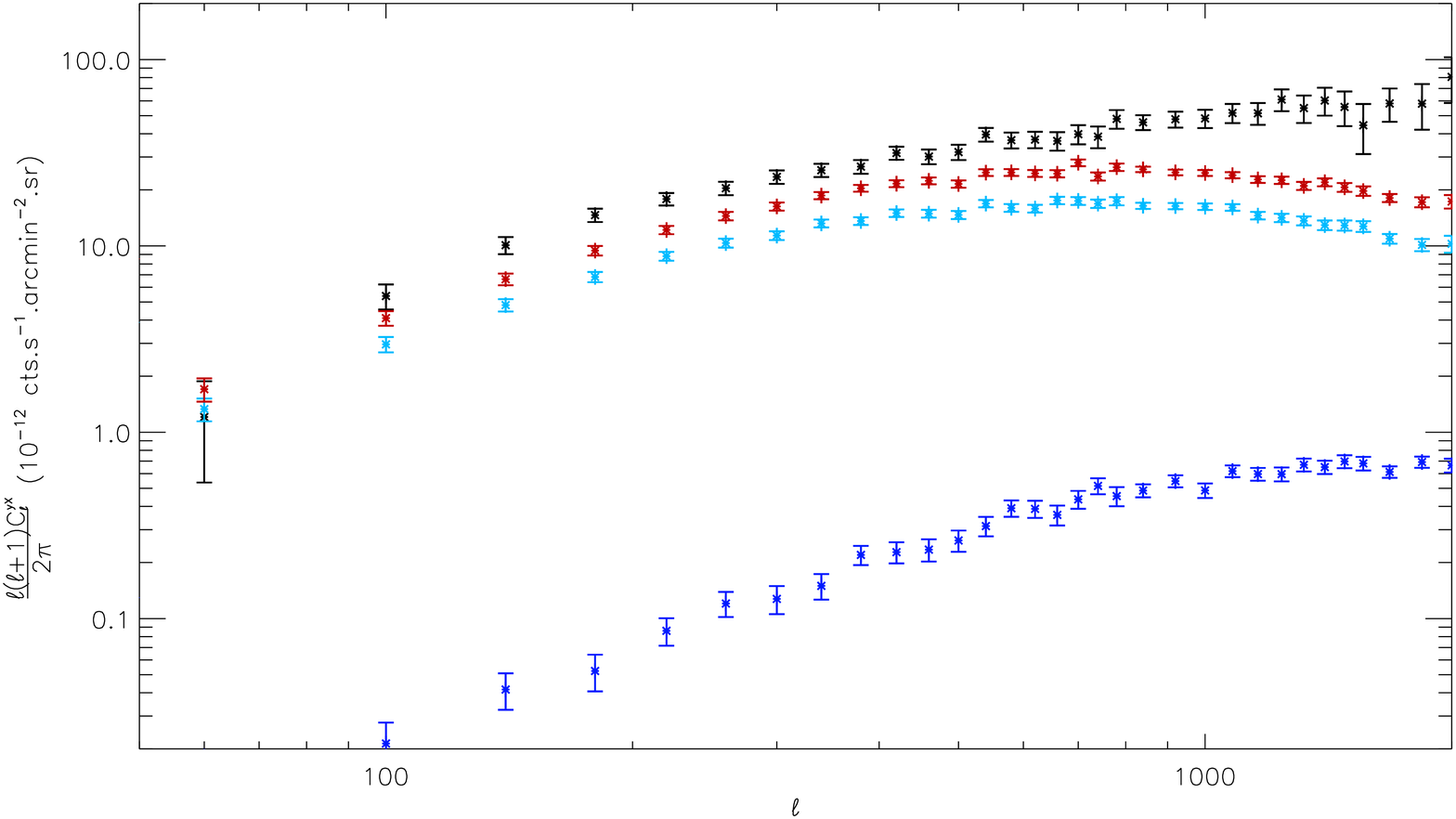}
\caption{In black we present the cross-correlation power spectrum between the RASS in the 0.5-2.0 keV energy range map and the MILCA $y$-map computed from {\it Planck} data, in red the contribution from all clusters in the MCXC catalogue \citep{pif11}, in light blue the contribution from all known clusters in the PSZ catalogue \citep{PlanckPSZ}, and in dark blue the contribution from the {\it Planck} cosmo sample used in \citep{PlanckSZC}.}
\label{crosscontrib}
\end{center}
\end{figure}

We also estimated the contribution of known tSZ and X-ray galaxy clusters to the tSZ-X cross-power spectrum. To do so, we built several tSZ maps for different subsamples of galaxy clusters. 
We considered the {\it Planck} SZ catalogue \citep{planckESZ,PlanckPSZ}, the {\it Planck} cosmo sample \citep{PlanckSZC}, and the MCXC
\citep{pif11}.\\
For the {\it Planck} SZ catalogue we used confirmed galaxy clusters (861 clusters) with fluxes $Y_{500}$ and radius $R_{500}$ estimated from 2D likelihoods provided in \citet{PlanckPSZ} and a universal pressure profile \citep{arn10}.
For X-rays clusters, we predicted the tSZ flux assuming a scaling relation between $Y_{500}$ and $L_{500}$ from \citet{planckSL} and X-ray deduced values for $R_{500}$.\\
We projected each cluster on an oversampled grid with a pixel size of 0.1$\times \theta_{500}$ (e.g. drizzling) to avoid flux loss during the projection.
Then we convolved the oversampled map with a beam FWHM of 10 arc minutes.
We reprojected the oversampled map on a HEALpix full-sky map with 1.7 arc minute pixels using a nearest-neighbour interpolation.\\

\begin{figure*}[!th]
\begin{center}
\includegraphics[scale=0.3]{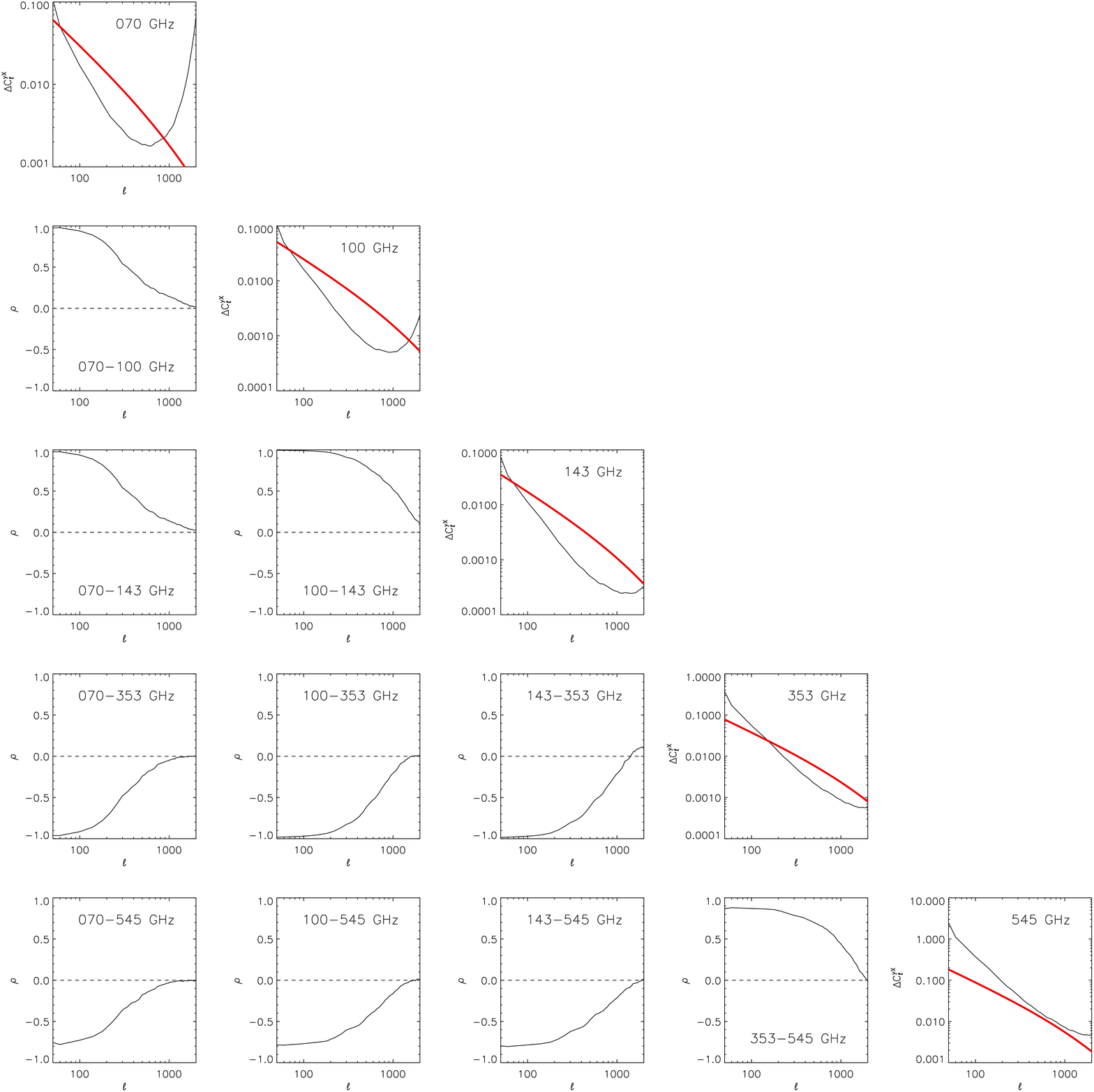}
\caption{Diagonal figures show in black the uncertainty level for $C_\ell^{\nu ,RASS}$ in units of K$_{\rm CMB}.$cts.s$^{-1}$.armin$^{-2}$.sr for the cleaned power spectra between the RASS map in the 0.5-2.0 keV energy range and the {\it Planck} maps at 70, 100, 143, 353, and 545 GHz, using a binning of $\Delta \ell = 40$, In red we show the tSZ-Xray cross-power spectrum for the model. The off diagonal figures show the correlation factor $\rho$ between $C_\ell^{yx}$ at frequencies $\nu$ and $\nu'$. See Sect.~\ref{secerr} for details on computing uncertainties and their covariance.}
\label{crosscovmat}
\end{center}
\end{figure*}

Then, we computed the angular cross-power spectrum between the tSZ template maps and the ROSAT full-sky count-rate map. Uncertainties were computed following the approach presented in in Sect.~\ref{secerr}. \\
In Fig.~\ref{crosscontrib}, we present the derived angular power spectra for each sample of clusters. The known X-ray clusters from the MCXC contribute about 75\% of the total power at low-$\ell$  ($\ell< 800$).  Confirmed {\it Planck} tSZ clusters contribute about 50\% of the power of the tSZ-X spectrum at low-$\ell$ ($\ell< 800$). The {\it Planck} cosmo sample contributes about 1\% at low-$\ell$.\\ 
Known clusters contribute essentially at low-$\ell$. At high-$\ell$, the tSZ-X spectrum presents contribution from low-mass and high-$z$ galaxy clusters. 
{\it Planck} -confirmed clusters and MCXC clusters only contribute 20\% and 40\% of the total power at $\ell = 2000$, as expected considering the contribution from undetected galaxy clusters.\\
These spectra demonstrate that the full tSZ-X cross-power spectrum contains additional informations on cosmology relevant to tSZ number count studies.

\section{Estimating bias and uncertainties}

We discuss in detail the main sources of uncertainties and biases in the measurement of the tSZ-X cross-power spectrum.

\subsection{Statistical uncertainties}
\label{secerr}

The cleaned angular cross-power spectra are constructed using a linear combination of three cross-power spectra, $C_\ell^{\nu,{\rm RASS}}$, at different frequencies, see Sect. \ref{secperfreq}. Consequently, we need to estimate variances and covariances of these spectra.
The variance, ${\rm var}(\nu)$, of $C_\ell^{\nu,{\rm RASS}}$ can be expressed as
\begin{equation}
{\rm var}(\nu) = \frac{1}{(2 \ell + 1) f_{\rm sky}} \left[ \left(C_\ell^{\nu,{\rm RASS}}\right)^2 +  C_\ell^{\nu,\nu}C_\ell^{{\rm RASS},{\rm RASS}}  \right],
\end{equation}
and the covariance, ${\rm cov}(\nu,\nu')$, between $C_\ell^{\nu,{\rm RASS}}$ and $C_\ell^{\nu',{\rm RASS}}$ reads
\begin{equation}
{\rm cov}(\nu,\nu') = \frac{1}{(2 \ell + 1) f_{\rm sky}} \left[ C_\ell^{\nu,{\rm RASS}} C_\ell^{\nu',{\rm RASS}} +  C_\ell^{\nu,\nu'}C_\ell^{{\rm RASS},{\rm RASS}}  \right].
\end{equation}
Then, the variance of $\widehat{C}_\ell^{\nu,{\rm RASS}}$ can be computed as 
\begin{align}
{\rm var}(\widehat{\nu}) =& \, {\rm var}(\nu) + {\rm var}(217) +  \left(\rho_{857,\nu} - \rho_{857,217} \right)^2{\rm var}(857) \nonumber \\
& - 2 \left(\rho_{857,\nu} - \rho_{857,217} \right) \left[ {\rm cov}(\nu,857)- 2 {\rm cov}(217,857) \right] \nonumber \\
& - 2 {\rm cov}(\nu,217).
\label{eqerrtx}
\end{align}
It can be useful to combine constraints from the different cleaned power spectra. 
To do so, we computed the covariance between $\widehat{C}_\ell^{\nu,{\rm RASS}}$ and $\widehat{C}_\ell^{\nu',{\rm RASS}}$ as

\begin{align}
{\rm cov}(\widehat{\nu},\widehat{\nu'}) =&\, {\rm cov}(\nu,\nu') + {\rm var}(217) \nonumber \\
&+  \left(\rho_{857,\nu} - \rho_{857,217} \right)\left(\rho_{857,\nu'} - \rho_{857,217} \right){\rm var}(857) \nonumber \\
& - {\rm cov}(\nu,217) - {\rm cov}(\nu',217) \nonumber \\
& - \left(\rho_{857,\nu} - \rho_{857,217} \right) \left[ {\rm cov}(\nu,857) - {\rm cov}(217,857) \right] \nonumber \\
& - \left(\rho_{857,\nu'} - \rho_{857,217} \right) \left[( {\rm cov}(\nu',857)- {\rm cov}(217,857) \right]. \nonumber \\
\label{eqcovtx}
\end{align}
We propagated the whole covariance matrix through binning and mask deconvolution processes.
In Fig.~\ref{crosscovmat}, we present the uncertainties on $\widehat{C}_\ell^{\nu,{\rm RASS}}$ and the correlation factor between $\widehat{C}_\ell^{\nu,{\rm RASS}}$ and $\widehat{C}_\ell^{\nu',{\rm RASS}}$ for the cleaned cross-power spectra at 70, 100, 143, 353, and 545 GHz.\\
In this figure and in the diagonal plots, we observe that at low frequencies the uncertainty level increases with the multipole. 
This is produced by the deconvolution from the instrumental resolution.
We observe a flattening in the uncertainty level at high-$\ell$ for the highest frequencies. This flattening indicates that the signal is dominated by the instrumental noise.\\
At low-$\ell$ the correlation factors are close to 1 or -1, indicating an almost total correlation between uncertainties below $\ell = 100$ at 70, 100, 143, and 353 GHz. 
This indicates that the uncertainties at these multipole values are dominated by cosmic variance of the tSZ-X cross-spectrum.\\
The instrumental noise domination at high-$\ell$ also appears in the correlation factor between frequencies. When the noise dominates the uncertainty budget, these coefficients become zero.\\

For a tSZ-X power spectrum estimated from a $y$-map, the variance, ${\rm var}(y)$, can be directly estimated as
\begin{equation}
{\rm var}(y) = \frac{1}{(2 \ell + 1) f_{\rm sky}} \left[ \left(C_\ell^{y,{\rm RASS}}\right)^2 +  C_\ell^{y,y}C_\ell^{{\rm RASS},{\rm RASS}}  \right].
\label{eqerryx}
\end{equation}
The MILCA method is tailored to minimize instrumental noise and the variance of other astrophysical emissions, therefore we derive lower uncertainties (Fig.~\ref{crossybestfit}) than using the previous approach (Fig.~\ref{crossfreqbestfit}).\\
We also probate these uncertainty through the binning and mask deconvolution processes.

\begin{figure}[!th]
\begin{center}
\includegraphics[scale=0.2]{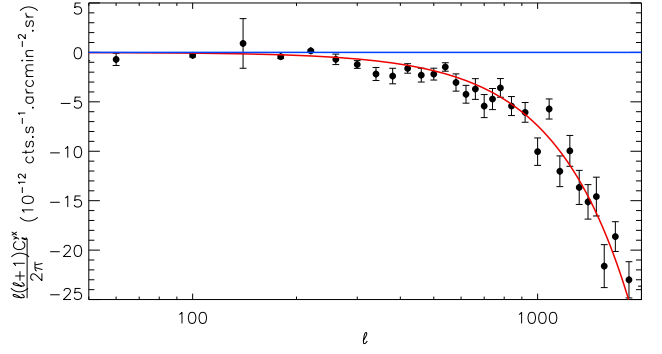}
\caption{Estimating the AGN contamination to the MILCA-RASS angular cross-power spectrum using NVSS and SUMSS catalogues as tracers of radio-loud AGNs. Black samples show our estimates, the solid blue line shows the 0 level, and the solid red line shows the best-fit of this contamination.}
\label{crossyagn}
\end{center}
\end{figure}

\begin{figure*}[!th]
\begin{center}
\includegraphics[scale=0.2]{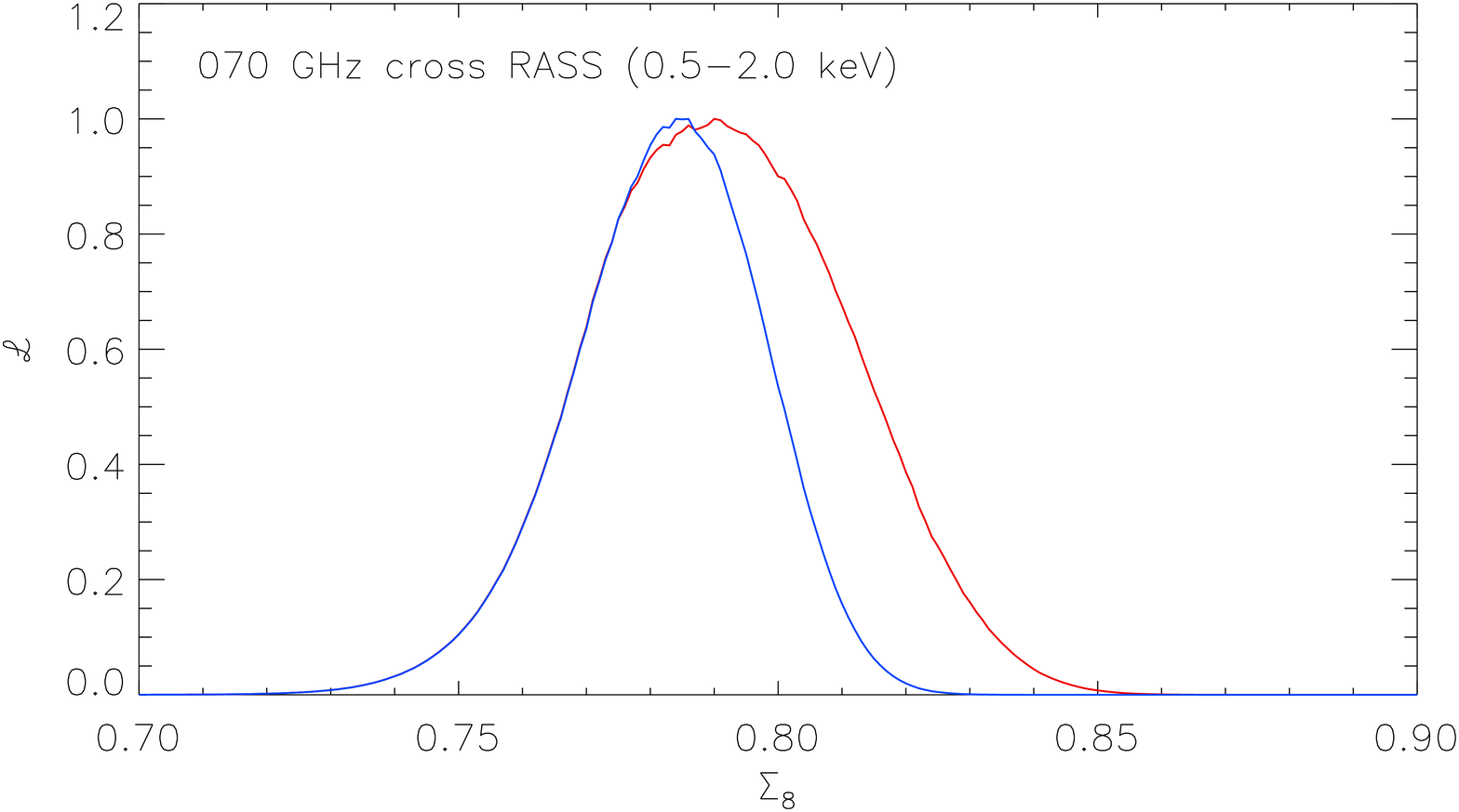}
\includegraphics[scale=0.2]{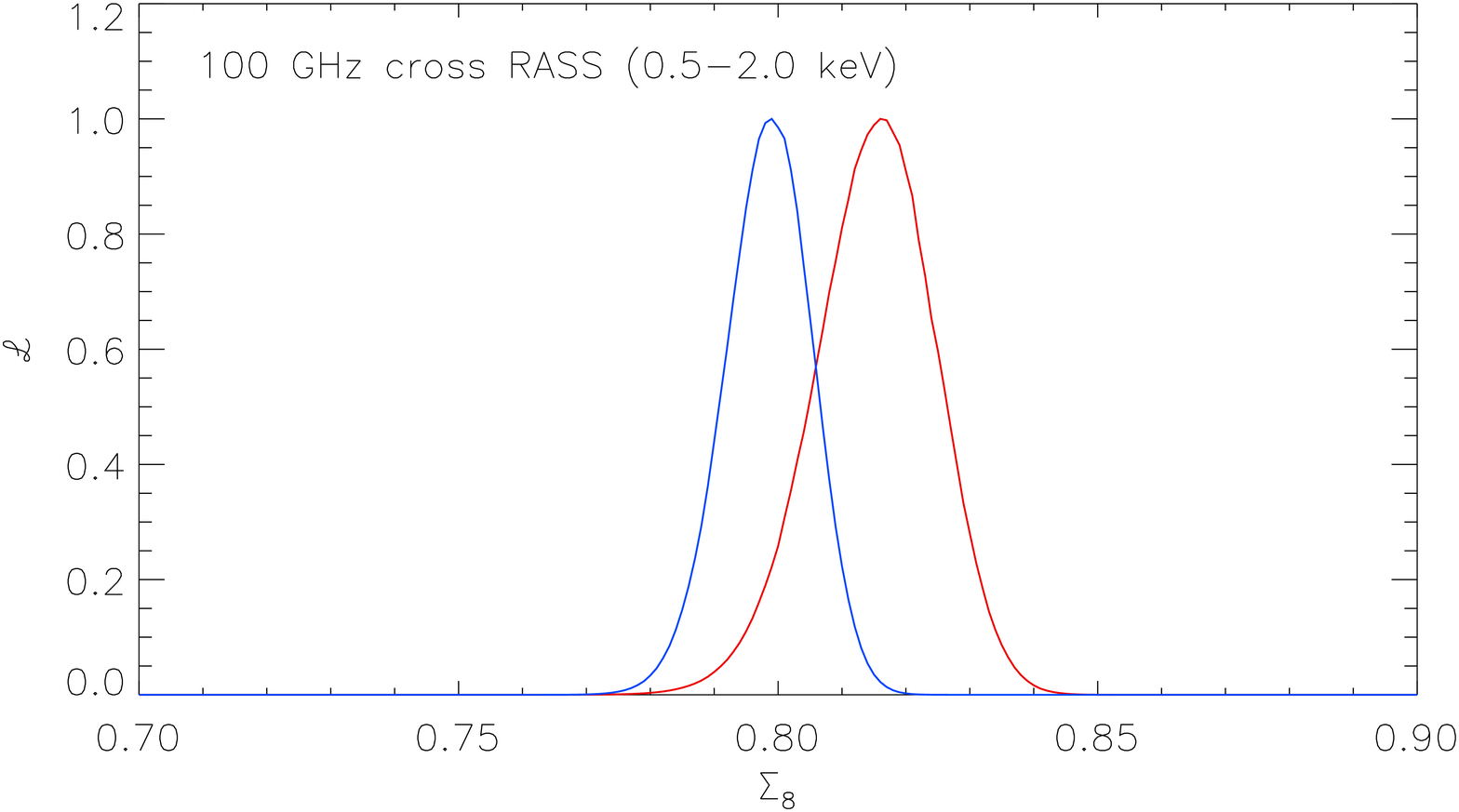}
\includegraphics[scale=0.2]{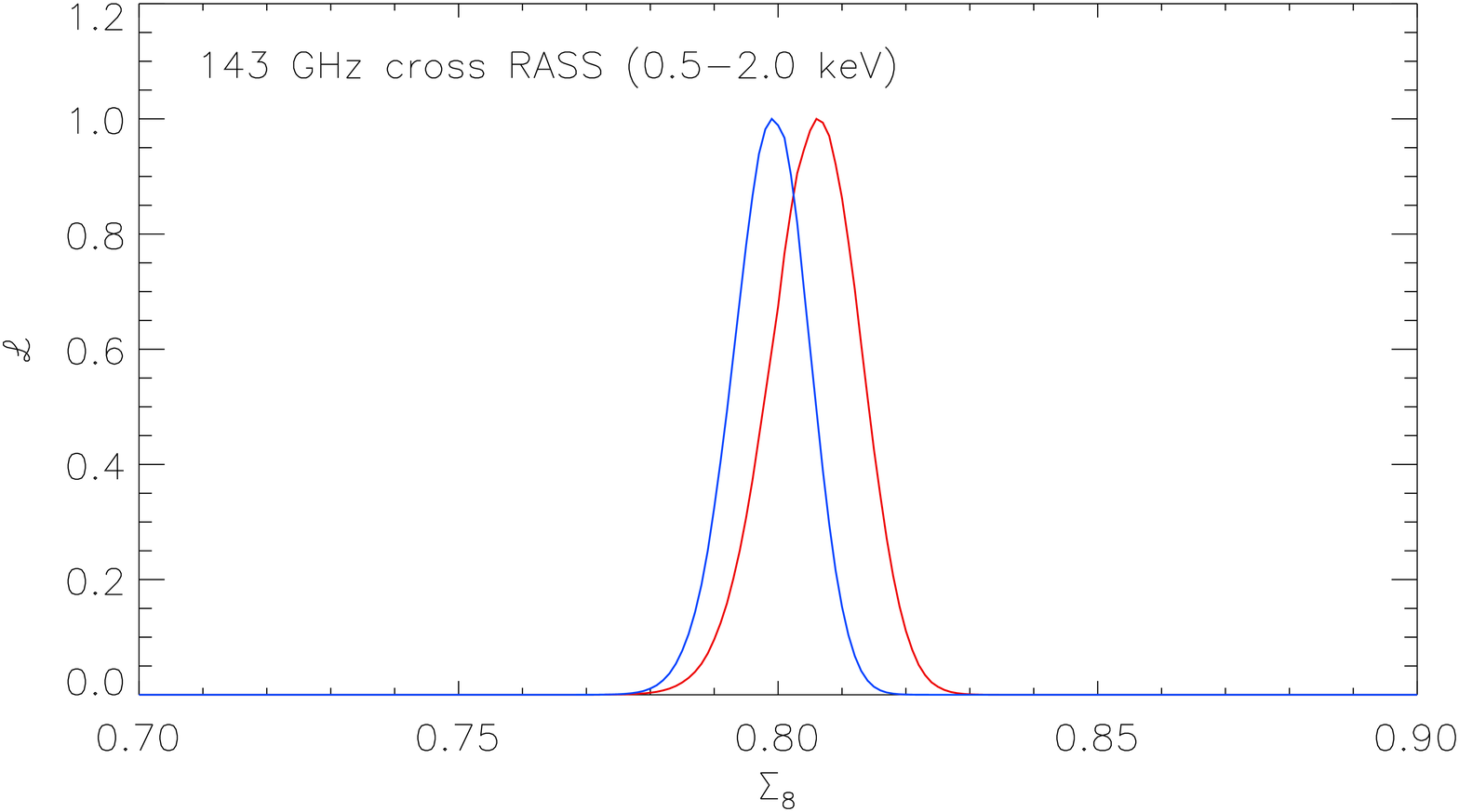}
\includegraphics[scale=0.2]{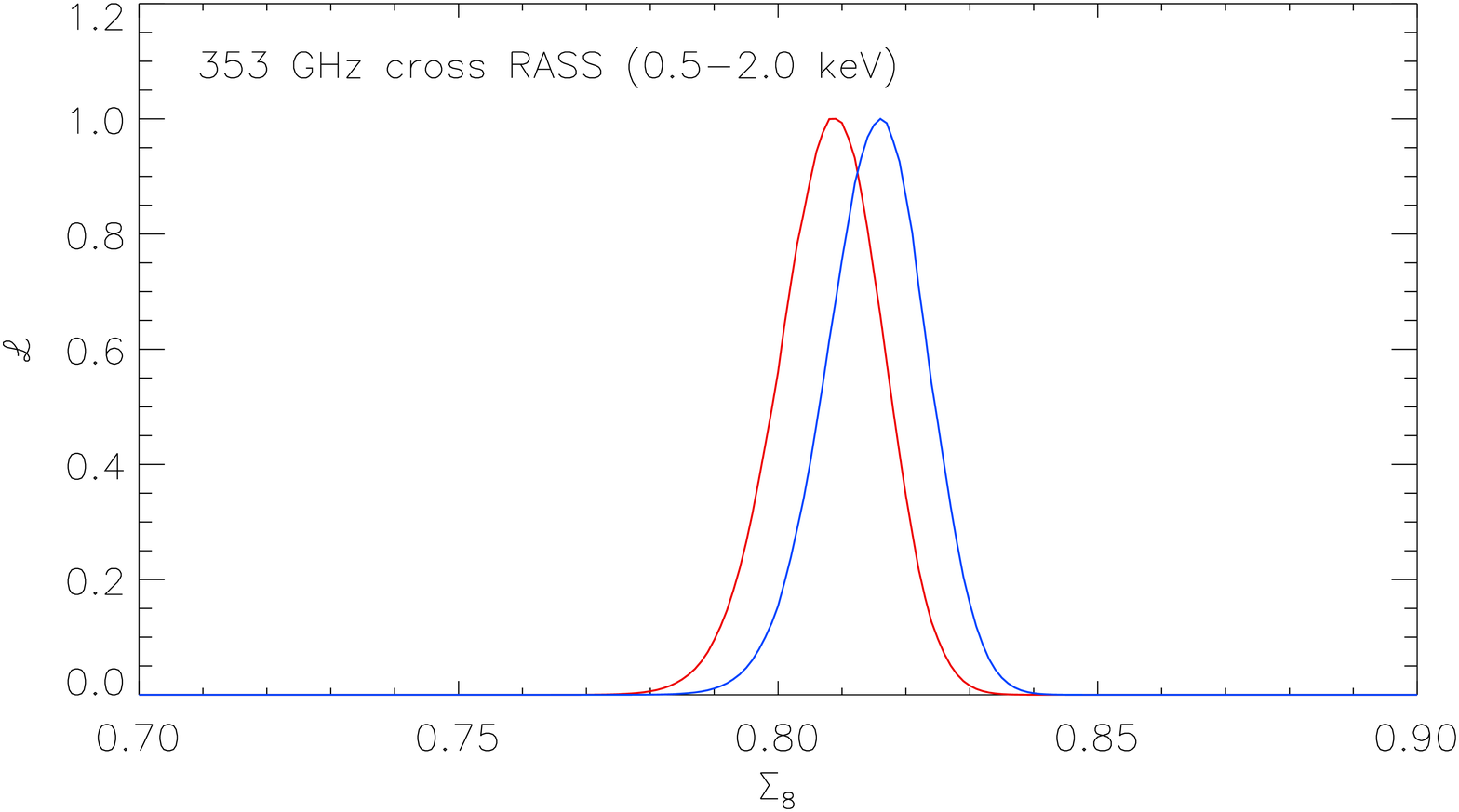}
\includegraphics[scale=0.2]{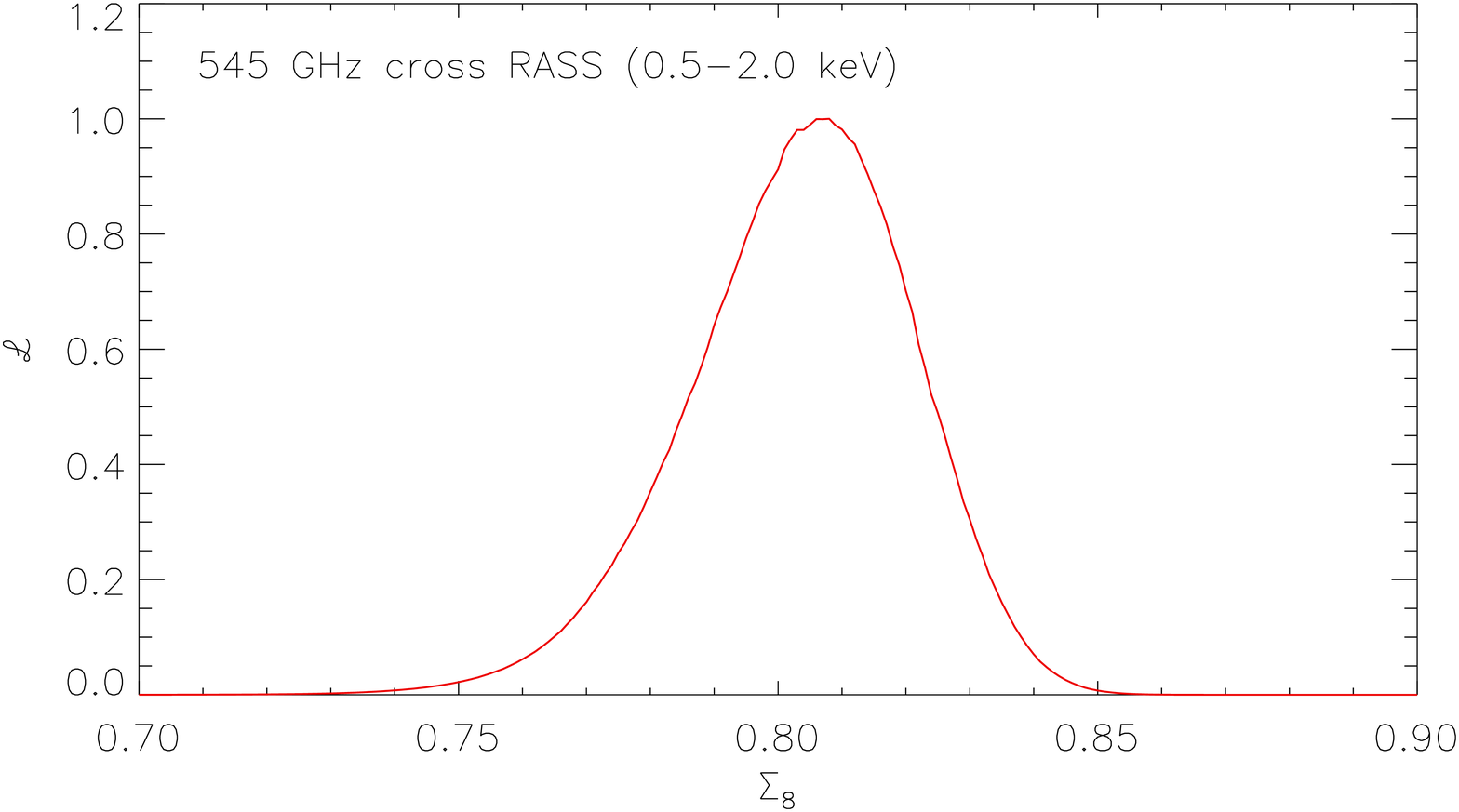}
\caption{From left to right and top to bottom: likelihood functions of $\Sigma_8$ estimated from the angular cross-power spectra at 70, 100, 143, 353, and 545 GHz, respectively. This adjustment accounts for AGNs and CIB contaminations. Considering modelling from Eq.~\ref{modsimp} in blue and from Eq.~\ref{modtot} in red.}
\label{likefreq1dmod}
\end{center}
\end{figure*}

\subsection{Bias sources}
There are several possible sources of bias in the tSZ-X cross-power spectrum.
The thermal dust emission from the Milky Way can produce bias
because it is correlated with the $n_{(H)}$ absorption of the X-rays.
By masking the galactic plane and considering X-rays in the range [0.5,2.0] keV, we ensure that this bias is under control \citep[see][for absorption cross-sections]{mor83}. 
We have tested several galactic cuts (20, 40, and 60\%) and found no significant variations in our measure (below 0.2 $\sigma$ on the tSZ-X cross-power spectra). This bias can thus be safely neglected.\\

There are also sources of biases that are correlated with the cluster spatial distribution over the sky. 
Indeed, all extra-galactic emissions are correlated with the matter distribution.
On the X-ray side, we have mainly two components, the X-ray emission from clusters and from AGNs (noted AGN$_{\rm X}$ hereafter).
On the microwave side, we have the tSZ effect, the radio emission from AGNs (noted AGN$_{\rm R}$ hereafter), and the cosmic infra-red background (CIB).
Consequently, we have five cross-spectra from extra-galactic objects that can bias the measurement.\\
The extrapolated radio emission from AGNs is weak at {\it Planck} frequencies \citep[e.g.][]{PlanckPSZ}, thus it does not produce a significant bias. 
In addition, the small number of un-masked high-flux X-ray AGNs prevents a significant contamination in terms of tSZ-AGN$_{\rm X}$ correlation.
As a consequence, we considered that tSZ-AGN$_{\rm X}$, AGN$_{\rm R}$-X, and CIB-AGN$_{\rm X}$ cross-correlations are weaker than
the AGN$_{\rm R}$-AGN$_{\rm X}$ correlations. 
This leaves the tSZ-X, AGN$_{\rm R}$-AGN$_{\rm X}$, and CIB-X correlations as potentially significant contributions for the microwave-X correlation.

For the multi-frequency estimation of the tSZ-X correlation (see Sect.~\ref{secperfreq}), this sources of bias can be separated from the tSZ-X signal using their SED.
However, this is not possible in the $y$-map approach (see Sect.~\ref{secymap}).
To estimate the AGN-AGN level of contamination in the tSZ-X angular cross-power obtained from the reprojected RASS full sky map and the $y$-map,
we assumed that radio catalogues are accurate tracers of the AGN population that may bias our measurement.\\

The NRAO VLA Sky Survey (NVSS)  \citep{con98} is a 1.4 GHz continuum 
survey covering the entire sky north of Dec$~ > -40^\circ$. The 
associated catalogue of 
discrete sources contains over 1.8 million radio sources. South of 
Dec$~ < -30^\circ$ and at galactic latitudes $|b| > 10^\circ$,
the Sydney University Molonglo Sky Survey (SUMSS)  \citep{mau03,mau08} 
is a 843 MHz continuum survey also providing a radio source catalogue. 
SUMSS and NVSS have similar sensitivities and angular resolutions, and
combined, they cover the whole sky.\\ 
We extrapolated the SUMSS sources at 1.4 GHz assuming a spectral index of -1 in intensity units.
Then, we built a map of NVSS sources weighted by their flux and filled the region Dec$~ > -40^\circ$ with extrapolated SUMSS sources.
We computed the cross-power spectra between the radio-sources map and the $y$-map, $C_\ell^{y,{\rm rad}}$ , and the RASS full-sky map, $C_\ell^{{\rm RASS},{\rm rad}}$.\\
Thus, we computed the estimate of the AGN-AGN contamination in the $C_\ell^{y,{\rm RASS}}$ spectrum as
\begin{equation}
C_\ell^{{\rm AGN}_{\rm R},{\rm AGN}_{\rm X}} = \frac{C_\ell^{y,{\rm rad}} C_\ell^{{\rm RASS},{\rm rad}}}{C_\ell^{{\rm rad},{\rm rad}}}.
\end{equation}

In Fig.~\ref{crossyagn} we present the derived $C_\ell^{{\rm AGN}_{\rm R},{\rm AGN}_{\rm X}}$ spectrum. We fitted $C_\ell^{{\rm AGN}_{\rm R},{\rm AGN}_{\rm X}}$ with a constant considering that AGN are point-like sources. Neglecting their clustering\footnote{AGN clustering can be safely neglected, the AGN power over the sky is dominated by a small number of objects. Similarly to what we observed for galaxy clusters, the clustering term is smaller
than the Poissonian term.}, we derive $C_\ell^{{\rm AGN}_{\rm R},{\rm AGN}_{\rm X}} = (-46.6 \pm 3.1)\, 10^{-18}$~cts.s$^{-1}$.arcmin$^{2}$.sr (red line in Fig.~\ref{crossyagn}).\\
Finally, we corrected the $y$-RASS cross-spectrum for this bias and propagated the related uncertainties.\\

\begin{table*}
\centering
\caption{Best-fitting values for $\Sigma_8$ for different data
subsets. From left to right: mean value of $\Sigma_8$, value of $\Sigma_8$ at the maximum likelihood, standard deviation of $\Sigma_8$, 68\% interval, systematic uncertainties, and $\chi^2$ per degree of freedom.}
\begin{tabular}{|c|c|c|c|c|c|c|c|}
\hline
Dataset & Model & $<\Sigma_8>$ & $\Sigma_{8,max}$ & $\sqrt{<{\Sigma_8}^2> - <{\Sigma_8}>^2}$ & 68\% interval & $\Delta$ Syst & $\chi^2_{\rm ndf}$ \\
\hline
X-rays $\times$ 70-545 GHz & Eq.~\ref{modtot} & 0.804 & 0.805 & 0.006 & 0.800 - 0.809 & 0.001 & 1.22 \\
\hline
X-rays $\times$ 70 GHz & Eq.~\ref{modsimp} & 0.782 & 0.784 & 0.015 & 0.770 - 0.798 & 0.020 & 1.51 \\
X-rays $\times$ 100 GHz & Eq.~\ref{modsimp} & 0.798 & 0.799 & 0.007 & 0.793 - 0.805 & 0.010 & 0.94\\
X-rays $\times$ 143 GHz & Eq.~\ref{modsimp} & 0.799 & 0.799 & 0.006 & 0.794 - 0.804 & 0.010 & 1.24 \\
X-rays $\times$ 353 GHz & Eq.~\ref{modsimp} & 0.815 & 0.816 & 0.008 & 0.809 - 0.823 & 0.015 & 0.61 \\
\hline
X-rays $\times$ 70 GHz & Eq.~\ref{modtot} & 0.791 & 0.790 & 0.020 & 0.771 - 0.809 & 0.001 & 1.51 \\
X-rays $\times$ 100 GHz & Eq.~\ref{modtot} & 0.815 & 0.816 & 0.010 & 0.807 - 0.824 & 0.001 & 0.85\\
X-rays $\times$ 143 GHz & Eq.~\ref{modtot} & 0.806 & 0.806 & 0.007 & 0.800 - 0.812 & 0.001 & 1.20 \\
X-rays $\times$ 353 GHz & Eq.~\ref{modtot} & 0.808 & 0.809 & 0.008 & 0.801 - 0.816 & 0.001 & 0.61 \\
X-rays $\times$ 545 GHz & Eq.~\ref{modtot} & 0.803 & 0.808 & 0.018 & 0.790 - 0.822 & 0.001 & 1.35 \\
\hline
X-rays $\times$ $y$-map & Eq.~\ref{mody} & 0.804 & 0.804 & 0.003 & 0.801 - 0.806 & 0.002 & 1.15 \\
\hline
\end{tabular}
\label{fit}
\end{table*}

\label{secbias}

\section{Constraints}
\label{seccosmo}

We explored three approaches to set cosmological constraints using the measurement of the tSZ-X cross-power spectrum. First, we considered individual frequencies, neglecting all sources of bias. Then, we considered the constraint from each frequency considering a multi-frequency adjustment for the biases. Finally, we considered the cross-spectrum with the tSZ $y$-map. In the following, we fit the data considering $\ell$ from 50 to 2000. We computed the likelihood functions assuming Gaussian uncertainties over a grid for $\Omega_m$, $\sigma_8$, and $H_0$, considering fixed values for other parameters. Then, we marginalized the likelihood to express it as a function of the degeneracy relation between parameters, $\Sigma_8$.
For cosmological constraints, we used fixed values of $a_{\rm x}$ and $a_{\rm sz}$ and propagated the uncertainties on these values to $\Sigma_8$.
In the following, uncertainties are given for 68\% confidence level.
We verified that we derived compatible results within the error bars using different galactic masks (see Sect.~\ref{secdata}). This demonstrates that our results are not significantly affected by galactic foreground contamination.

\subsection{Cosmological constraints per frequency channel}

Considering a single-frequency approach, we estimated $\Sigma_8$ for each frequency individually, excluding 545 GHz, which is contaminated. We assumed the model
\begin{equation}
\widehat{C}_\ell^{\nu,{\rm RASS}} = g(\nu)C_\ell^{yx}(\Sigma_8^{\nu}).
\label{modsimp}
\end{equation}
The derived likelihoods are presented in Fig.~\ref{likefreq1dmod} and the best-fitting values are summarized in Table.~\ref{fit}. 
Our best-fitting values for $\Sigma_8^{\nu}$ increase with frequency
with this simple modelling. 
This behaviour is produced by the contamination of AGN and CIB. The radio-loud AGNs produce an excess of correlation at low frequency that compensates for the anti-correlation between tSZ and X-ray emission.
At high frequency, the CIB-X correlation produces an excess of correlation that biases the correlation between tSZ and X-ray emission.
The values deduced from spectra at 100 GHz agree well with those
at 143 GHz, which present the lowest bias level.\\

\subsection{Cosmological constraints from the multi-frequency approach}
\label{fitall}

To prevent contamination produced by AGNs and CIB in the fit, we modelled the measured spectra, $\widehat{C}_\ell^{\nu,{\rm RASS}}$, as follows:
\begin{equation}
\widehat{C}_\ell^{\nu,{\rm RASS}} = \left[g(\nu) + A_{\rm CIB}f_{\rm CIB}(\nu)\right]C_\ell^{yx}(\Sigma_8^{\nu}) + A_{\rm rad} f_{\rm rad}(\nu),
\label{modtot}
\end{equation}
where $f_{\rm CIB}$ \citep{gis00} and $f_{\rm rad}(\nu)$ \citep{planckRS} are fixed SED for CIB and radio-loud AGN contaminations. \\
Given the large uncertainties at 545 GHz for the tSZ-X cross-correlation measurement, the signal is not sensitive to the shape of the CIB-X correlation. Therefore and for simplicity, we assumed that the CIB-X angular cross-power spectrum has the same shape, with respect to $\ell$, as the tSZ-X cross-correlation.\\
We fitted for $A_{\rm rad}$, $A_{\rm CIB}$ and $\Sigma_8^{\nu}$ the amplitude of the radio and CIB contamination and the tSZ-X angular cross-power spectrum degenerate amplitude defined in Eq.~\ref{eqdep}. Parameters $A_{\rm rad}$ and $A_{\rm CIB}$ were fitted using the five frequencies 70, 100, 143, 353, and 545 GHz;  $\Sigma_8^{\nu}$ was fitted individually for each frequency.\\
The derived likelihoods are presented in Fig.~\ref{likefreq1dmod}. Best-fitting values are summarized in Table.~\ref{fit}. The $\Sigma_8$ values now agree well, including the value deduced from the 545 GHz spectra. There is no significant frequency dependence for $\Sigma_8$, indicating that our modelling accounts properly for contamination by other astrophysical components.\\ 

\begin{figure}[!th]
\begin{center}
\includegraphics[scale=0.2]{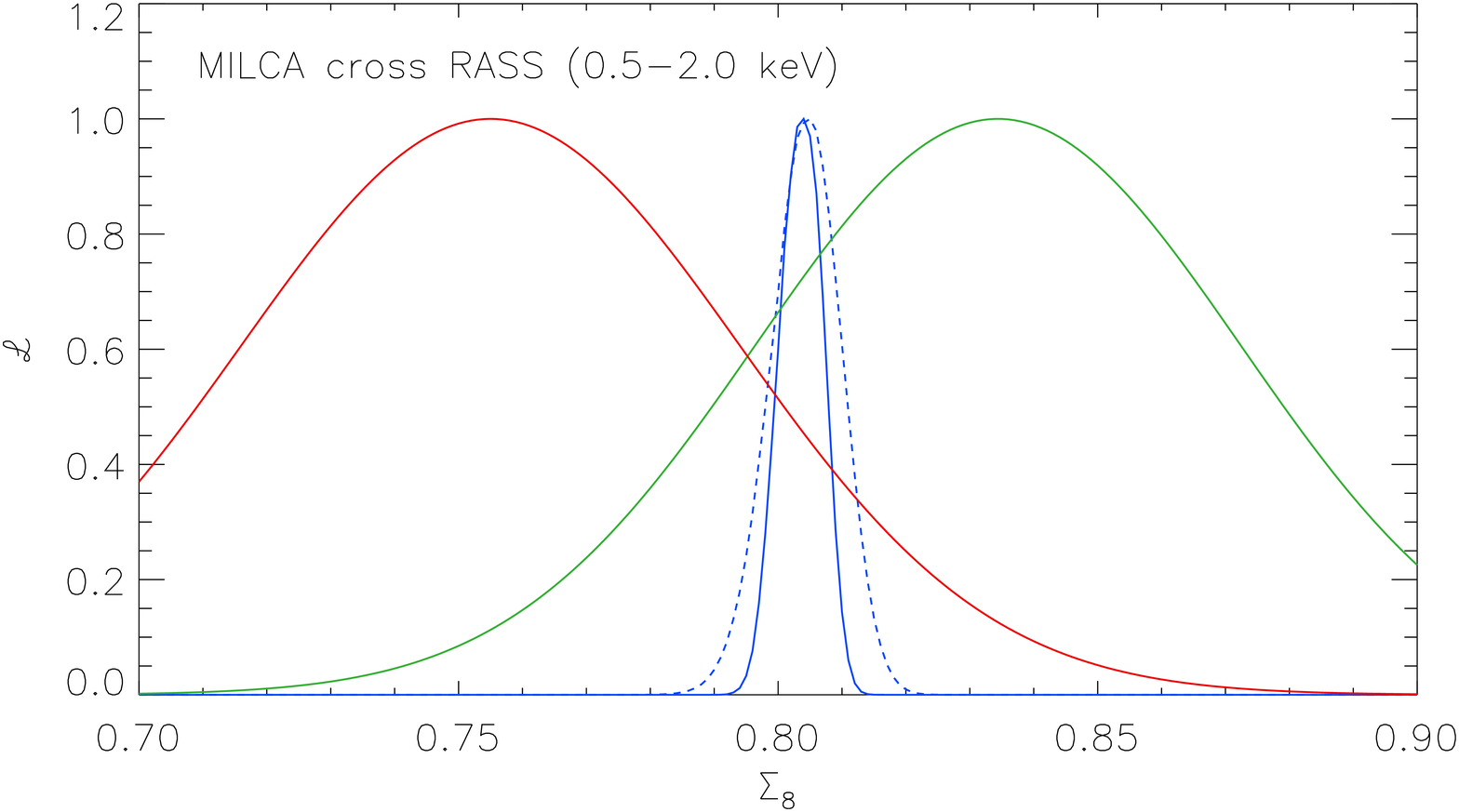}
\includegraphics[scale=0.2]{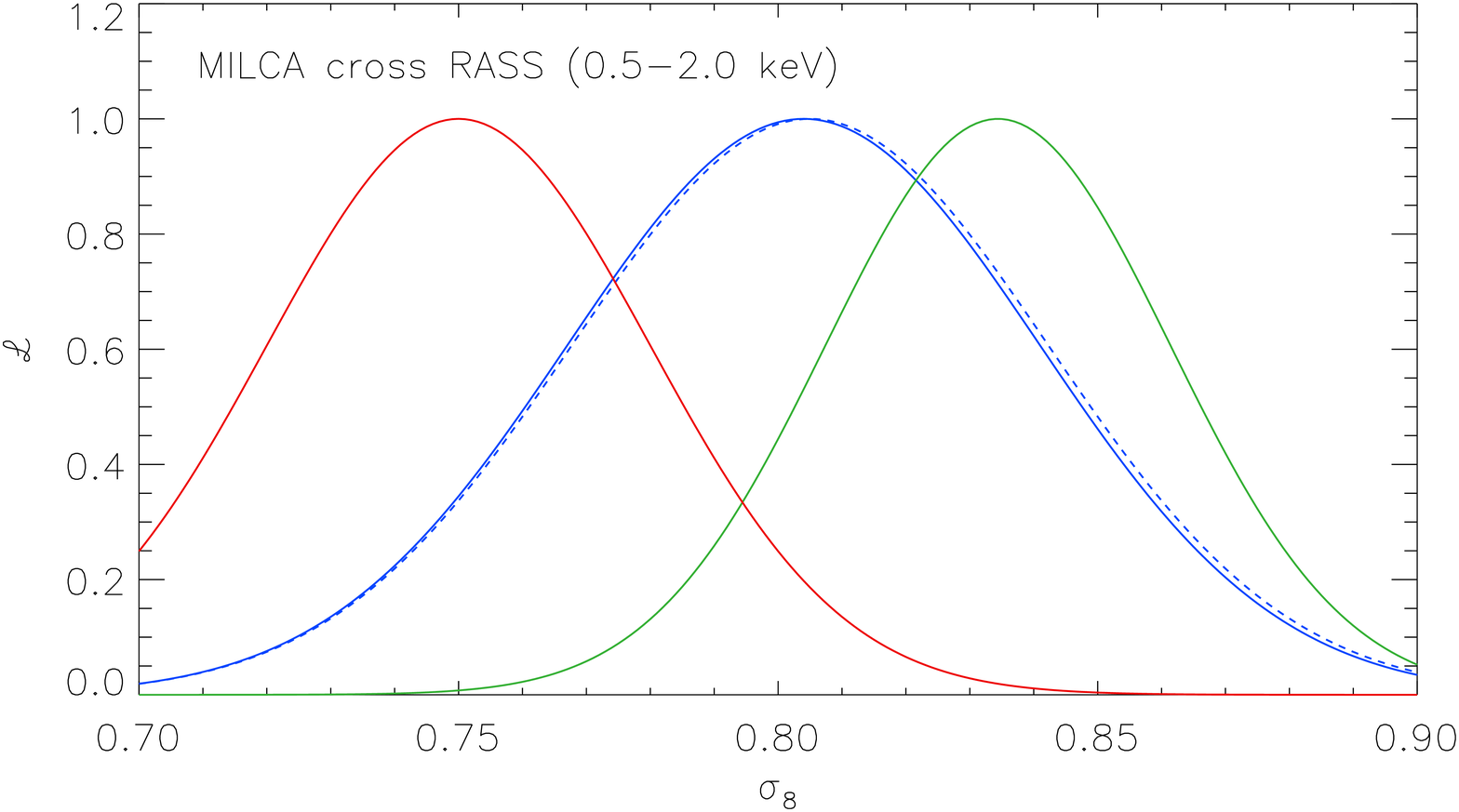}
\caption{Top panel: constraints on $\Sigma_8$. Bottom panel: associated constraints on $\sigma_8$ considering the modelling uncertainties.
The solid dark blue line depicts the likelihood functions estimated from the angular cross-power spectra between the MILCA and the ROSAT full-sky maps. The dashed dark blue line represents the
likelihood function estimated from the cleaned cross-power spectra at 70, 100, 143, 353, and 545 GHz. This adjustment accounts for AGN contamination. In red we show the constraints considering the cosmology from the cluster number count, in green the constraints considering the cosmology from {\it Planck}-CMB.}
\label{likexsz}
\end{center}
\end{figure}

Then, we adjusted $\Sigma_8$ using the model presented in Eq.~\ref{modtot} and all frequencies simultaneously.
Figure~\ref{likexsz} presents the likelihood function for $\Sigma_8$ and $\sigma_8$. We obtain a best-fit value of $\Sigma_8 = 0.804 \pm 0.006 \pm 0.001$, with statistical and systematic uncertainties, respectively, where systematic uncertainties account for residual contamination. \\
We estimated the expected value for $\Sigma_8$ considering the
cosmological constraints from analyses based on  CMB \citep{planckPAR} and cluster counts \citep{PlanckSZC}.
Cluster number counts and CMB power spectra present different degeneracies for cosmological parameters from those of tSZ-X cross-spectrum. This leads to large error bars when $\Sigma_8$ is estimated from cluster counts or CMB. Uncertainty levels derived assuming cluster number count and CMB cosmology are described in \citet{hur14a}.\\
We stress that these uncertainties are not representative of the uncertainties on $\sigma_8$.
We need to propagate the uncertainty on the modelling parameters to $\sigma_8$.
Consequently, the constraints on cosmological parameters read $\sigma_8 \left[\left(\frac{\Omega_m}{0.3175}\right)^{3.42} \left(\frac{H_0}{67}\right)^{2.36}\right]^{\frac{1}{8.12}} = 0.805 \pm 0.006 \pm 0.001 \pm 0.025,$ with statistical, systematic, and modeling uncertainties, respectively. The constraint on $\sigma_8$ gives $\sigma_8 = 0.805 \pm 0.037.$
We observe that the error budget is dominated by modelling uncertainties.

\subsection{Cosmological constraints from a $y$-map}

We estimated the level of CIB-X contamination in the $y$-RASS power spectrum by propagating the $A_{\rm CIB}$ amplitude through the MILCA linear weights. We deduce that this contamination is negligible. 
Indeed, the CIB only contaminates the tSZ-X correlation at high frequency, where the weights are smaller than at low frequencies.

To reproduce the $y$-RASS spectrum, we assumed the following modedling:
\begin{equation}
C_\ell^{y,{\rm RASS}} = C_\ell^{yx}(\Sigma_8) + A_{\rm rad},
\label{mody}
\end{equation}
where $A_{\rm rad}$ is estimated as presented in Sect.~\ref{secbias}.
 From this measurement, we obtain $\sigma_8 \left[\left(\frac{\Omega_m}{0.3175}\right)^{3.42} \left(\frac{H_0}{67}\right)^{2.36}\right]^{\frac{1}{8.12}} = 0.804 \pm 0.003 \pm 0.002 \pm 0.025$ and $\sigma_8 = 0.804 \pm 0.037$.\\

In Fig.~\ref{likexsz}, we present the likelihood for $\Sigma_8$ and $\sigma_8$ .
The constraints on $\Sigma_8$ derived from the tSZ-X power spectrum are consistent with predictions from cluster counts and CMB.
Moreover, the constraints derived from the $y$-map are consistent with constraints derived from the multi-frequency analysis presented in Sect.~\ref{fitall}.

\subsection{Constraints on scaling-law parameters}
\label{secclus}

\begin{figure}[!th]
\begin{center}
\includegraphics[scale=0.2]{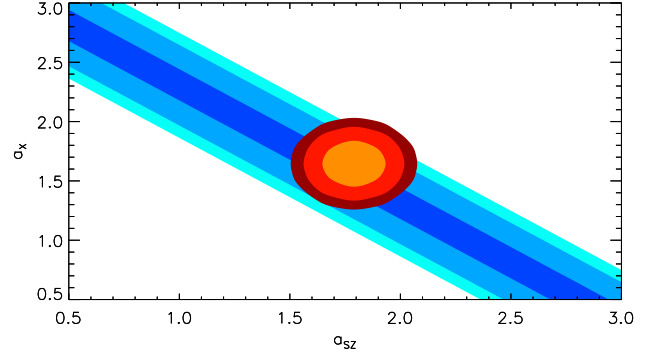}
\caption{Two-dimensional likelihood function for the scaling relation mass power-laws, $a_{\rm sz}$ and $a_{\rm x}$. In blue we present the 1, 2 and, 3 $\sigma$ confidence levels from our measurement of the tSZ-X cross-spectrum, in red the 1, 2 and 3 $\sigma$ confidence levels for our constraints on the scaling-law mass indices.}
\label{likexszind}
\end{center}
\end{figure}

In Fig.~\ref{likexszind}, we present the constraints on the scaling relation power-law indices from combined tSZ-X and tSZ-tSZ power spectra. 
We assumed a fixed pressure profile \citep[best-fit on a GNFW profile from][]{arn10} and polytropic index of 1.5 for galaxy clusters.\\
From the tSZ-Xray cross-spectrum, we derive $a_{\rm sz} + a_{\rm x} = 3.37 \pm 0.09$, which is to be compared with $a_{\rm sz} + a_{\rm x} =  3.43 \pm 0.15$ when considering values listed in Table.~\ref{tabscal}.
Figure~\ref{likexszind} shows that constraints from the tSZ-X spectrum alone are consistent with previous constraints \citep[see e.g.][]{pra09,planckSL} on $a_{\rm x}$ and $a_{\rm sz}$ .\\

To explore the possible evolution of the scaling relations with redshift, we assumed a power-law dependence of the form $(1+z)^{\beta}$. Then, we fitted for $\beta_{\rm sz} + \beta_{\rm x}$ the tSZ and X-ray evolution indices.
We derive $\beta_{\rm sz} + \beta_{\rm x} = 0.4^{+0.4}_{-0.5}$ from the tSZ-X angular cross-power spectrum.

We stress that the constraints on scaling-law indices are degenerate with the assumed density and temperature profiles for galaxy clusters \citep[as discussed in][]{hur14a}.

\section{Conclusion}
\label{concl}

We have performed the first detection of the tSZ-X angular cross-power spectrum directly estimated from X-ray count rate and tSZ signal.
We detected a clearly correlated emission between tSZ-map and X-ray count-rate map in the range $50<\ell < 2000$ at an overall significance of 28 $\sigma$.\\
We accounted for statistical systematics from residuals (mainly AGN and CIB) and for modelling uncertainties.\\

Using the amplitude of the tSZ-X angular cross-power spectrum, we derived constraints on cosmological parameters $\sigma_8 \left[\left(\frac{\Omega_m}{0.3175}\right)^{3.42} \left(\frac{H_0}{67}\right)^{2.36}\right]^{\frac{1}{8.12}} = 0.804 \pm 0.003 \pm 0.002 \pm 0.025$ with statistical, systematic, and modelling uncertainties.
These constraints read $\sigma_8 = 0.804 \pm 0.037$ if we consider $\Omega_m$ from {\it Planck}-CMB best-fit cosmology \citep{planckPAR}.
We tested the robustness of this result using an analysis based on multiple frequencies and on a tSZ-map.\\
With a mass bias of 0.2, this result is compatible with constraints
based on cluster count and CMB on $\sigma_8$ from \citet{PlanckSZC} and \citet{planckPAR}. This constraint is also consistent with the cross-correlation measured by \citet{haj13} between the tSZ emission and an X-ray sample of galaxy clusters and with the X-ray cluster number count \citep{vik09,boh14}. \\

Using the shape of the tSZ-X cross-spectrum, we were able to constrain the scaling-law mass and evolution indices of $L_{500}-M_{500}$ and $Y_{500}-M_{500}$.
We derived $a_{\rm sz} + a_{\rm x} = 3.37 \pm 0.09$ and $\beta_{\rm sz} + \beta_{\rm x} = 0.4^{+0.4}_{-0.5}$.
Consistently with previous results from \citet{rei11} for the X-ray scaling relation, we found that the redshift evolution of the scaling law is consistent with self-similar evolution.\\

\section*{Acknowledgements}
\thanks{The authors thanks A.beelen, M.Arnaud and G.Pratt for useful discussions.
We acknowledge the support of the French \emph{Agence Nationale de la Recherche} under grant ANR-11-BD56-015. 
This research has made use of the ROSAT all-sky survey data which have been processed at MPE. 
The development of {\it Planck} has been supported by: ESA; CNES and CNRS/INSU-IN2P3-INP (France); ASI, CNR, and INAF (Italy); NASA and DoE (USA); STFC and UKSA (UK); CSIC, MICINN and JA (Spain); Tekes, AoF and CSC (Finland); DLR and MPG (Germany); CSA (Canada); DTU Space (Denmark); SER/SSO (Switzerland); RCN (Norway); SFI (Ireland); FCT/MCTES (Portugal); and The development of {\it Planck} has been supported by: ESA; CNES and CNRS/INSU-IN2P3-INP (France); ASI, CNR, and INAF (Italy); NASA and DoE (USA); STFC and UKSA (UK); CSIC, MICINN and JA (Spain); Tekes, AoF and CSC (Finland); DLR and MPG (Germany); CSA (Canada); DTU Space (Denmark); SER/SSO (Switzerland); RCN (Norway); SFI (Ireland); FCT/MCTES (Portugal); and PRACE (EU).}

\bibliographystyle{aa}
\bibliography{powspec_xsz_mes}

\begin{thebibliography}{78}
\expandafter\ifx\csname natexlab\endcsname\relax\def\natexlab#1{#1}\fi

\bibitem[{{Abell} {et~al.}(1989){Abell}, {Corwin}, \& {Olowin}}]{abe89}
{Abell}, G.~O., {Corwin}, Jr., H.~G., \& {Olowin}, R.~P. 1989, \apjs, 70, 1

\bibitem[{{Afshordi} {et~al.}(2007){Afshordi}, {Lin}, {Nagai}, \&
  {Sanderson}}]{afs07}
{Afshordi}, N., {Lin}, Y.-T., {Nagai}, D., \& {Sanderson}, A.~J.~R. 2007,
  \mnras, 378, 293

\bibitem[{{Afshordi} {et~al.}(2005){Afshordi}, {Lin}, \& {Sanderson}}]{afs05}
{Afshordi}, N., {Lin}, Y.-T., \& {Sanderson}, A.~J.~R. 2005, \apj, 629, 1

\bibitem[{{Arnaud} {et~al.}(2010){Arnaud}, {Pratt}, {Piffaretti},
  {B{\"o}hringer}, {Croston}, \& {Pointecouteau}}]{arn10}
{Arnaud}, M., {Pratt}, G.~W., {Piffaretti}, R., {et~al.} 2010, \aap, 517, A92

\bibitem[{{Atrio-Barandela} {et~al.}(2008){Atrio-Barandela}, {Kashlinsky},
  {Kocevski}, \& {Ebeling}}]{atr08}
{Atrio-Barandela}, F., {Kashlinsky}, A., {Kocevski}, D., \& {Ebeling}, H. 2008,
  \apjl, 675, L57

\bibitem[{{Bennett} {et~al.}(2003){Bennett}, {Hill}, {Hinshaw}, {Nolta},
  {Odegard}, {Page}, {Spergel}, {Weiland}, {Wright}, {Halpern}, {Jarosik},
  {Kogut}, {Limon}, {Meyer}, {Tucker}, \& {Wollack}}]{ben03}
{Bennett}, C.~L., {Hill}, R.~S., {Hinshaw}, G., {et~al.} 2003, \apjs, 148, 97

\bibitem[{{Benson} {et~al.}(2013){Benson}, {de Haan}, {Dudley}, {Reichardt},
  {Aird}, {Andersson}, {Armstrong}, {Ashby}, {Bautz}, {Bayliss}, {Bazin},
  {Bleem}, {Brodwin}, {Carlstrom}, {Chang}, {Cho}, {Clocchiatti}, {Crawford},
  {Crites}, {Desai}, {Dobbs}, {Foley}, {Forman}, {George}, {Gladders},
  {Gonzalez}, {Halverson}, {Harrington}, {High}, {Holder}, {Holzapfel},
  {Hoover}, {Hrubes}, {Jones}, {Joy}, {Keisler}, {Knox}, {Lee}, {Leitch},
  {Liu}, {Lueker}, {Luong-Van}, {Mantz}, {Marrone}, {McDonald}, {McMahon},
  {Mehl}, {Meyer}, {Mocanu}, {Mohr}, {Montroy}, {Murray}, {Natoli}, {Padin},
  {Plagge}, {Pryke}, {Rest}, {Ruel}, {Ruhl}, {Saliwanchik}, {Saro}, {Sayre},
  {Schaffer}, {Shaw}, {Shirokoff}, {Song}, {Spieler}, {Stalder},
  {Staniszewski}, {Stark}, {Story}, {Stubbs}, {Suhada}, {van Engelen},
  {Vanderlinde}, {Vieira}, {Vikhlinin}, {Williamson}, {Zahn}, \&
  {Zenteno}}]{ben13}
{Benson}, B.~A., {de Haan}, T., {Dudley}, J.~P., {et~al.} 2013, \apj, 763, 147

\bibitem[{{Bleem} {et~al.}(2014){Bleem}, {Stalder}, {de Haan}, {Aird}, {Allen},
  {Applegate}, {Ashby}, {Bautz}, {Bayliss}, {Benson}, {Bocquet}, {Brodwin},
  {Carlstrom}, {Chang}, {Chiu}, {Cho}, {Clocchiatti}, {Crawford}, {Crites},
  {Desai}, {Dietrich}, {Dobbs}, {Foley}, {Forman}, {George}, {Gladders},
  {Gonzalez}, {Halverson}, {Hennig}, {Hoekstra}, {Holder}, {Holzapfel},
  {Hrubes}, {Jones}, {Keisler}, {Knox}, {Lee}, {Leitch}, {Liu}, {Lueker},
  {Luong-Van}, {Mantz}, {Marrone}, {McDonald}, {McMahon}, {Meyer}, {Mocanu},
  {Mohr}, {Murray}, {Padin}, {Pryke}, {Reichardt}, {Rest}, {Ruel}, {Ruhl},
  {Saliwanchik}, {Saro}, {Sayre}, {Schaffer}, {Schrabback}, {Shirokoff},
  {Song}, {Spieler}, {Stanford}, {Staniszewski}, {Stark}, {Story}, {Stubbs},
  {Vanderlinde}, {Vieira}, {Vikhlinin}, {Williamson}, {Zahn}, \&
  {Zenteno}}]{ble14}
{Bleem}, L.~E., {Stalder}, B., {de Haan}, T., {et~al.} 2014, ArXiv e-prints

\bibitem[{{Bobin} {et~al.}(2014){Bobin}, {Sureau}, {Starck}, {Rassat}, \&
  {Paykari}}]{bob14}
{Bobin}, J., {Sureau}, F., {Starck}, J.-L., {Rassat}, A., \& {Paykari}, P.
  2014, \aap, 563, A105

\bibitem[{{B{\"o}hringer} {et~al.}(2014){B{\"o}hringer}, {Chon}, \&
  {Collins}}]{boh14}
{B{\"o}hringer}, H., {Chon}, G., \& {Collins}, C.~A. 2014, ArXiv e-prints

\bibitem[{{B{\"o}hringer} {et~al.}(2001){B{\"o}hringer}, {Schuecker}, {Guzzo},
  {Collins}, {Voges}, {Schindler}, {Neumann}, {Cruddace}, {De Grandi},
  {Chincarini}, {Edge}, {MacGillivray}, \& {Shaver}}]{boh01}
{B{\"o}hringer}, H., {Schuecker}, P., {Guzzo}, L., {et~al.} 2001, \aap, 369,
  826

\bibitem[{{Bohringer} {et~al.}(2000){Bohringer}, {Voges}, {Huchra}, {McLean},
  {Giacconi}, {Rosati}, {Burg}, {Mader}, {Schuecker}, {Simic}, {Komossa},
  {Reiprich}, {Retzlaff}, \& {Trumper}}]{boh00}
{Bohringer}, H., {Voges}, W., {Huchra}, J.~P., {et~al.} 2000, VizieR Online
  Data Catalog, 212, 90435

\bibitem[{{Cole} \& {Kaiser}(1988)}]{col88}
{Cole}, S. \& {Kaiser}, N. 1988, \mnras, 233, 637

\bibitem[{{Condon} {et~al.}(1998){Condon}, {Cotton}, {Greisen}, {Yin},
  {Perley}, {Taylor}, \& {Broderick}}]{con98}
{Condon}, J.~J., {Cotton}, W.~D., {Greisen}, E.~W., {et~al.} 1998, \aj, 115,
  1693

\bibitem[{{Diego} \& {Majumdar}(2004)}]{die04}
{Diego}, J.~M. \& {Majumdar}, S. 2004, \mnras, 352, 993

\bibitem[{{Diego} \& {Partridge}(2010)}]{die10}
{Diego}, J.~M. \& {Partridge}, B. 2010, \mnras, 402, 1179

\bibitem[{{Diego} {et~al.}(2003){Diego}, {Silk}, \& {Sliwa}}]{die03}
{Diego}, J.~M., {Silk}, J., \& {Sliwa}, W. 2003, \mnras, 346, 940

\bibitem[{{Dunkley} {et~al.}(2011){Dunkley}, {Hlozek}, {Sievers}, {Acquaviva},
  {Ade}, {Aguirre}, {Amiri}, {Appel}, {Barrientos}, {Battistelli}, {Bond},
  {Brown}, {Burger}, {Chervenak}, {Das}, {Devlin}, {Dicker}, {Bertrand
  Doriese}, {D{\"u}nner}, {Essinger-Hileman}, {Fisher}, {Fowler}, {Hajian},
  {Halpern}, {Hasselfield}, {Hern{\'a}ndez-Monteagudo}, {Hilton}, {Hilton},
  {Hincks}, {Huffenberger}, {Hughes}, {Hughes}, {Infante}, {Irwin}, {Juin},
  {Kaul}, {Klein}, {Kosowsky}, {Lau}, {Limon}, {Lin}, {Lupton}, {Marriage},
  {Marsden}, {Mauskopf}, {Menanteau}, {Moodley}, {Moseley}, {Netterfield},
  {Niemack}, {Nolta}, {Page}, {Parker}, {Partridge}, {Reid}, {Sehgal},
  {Sherwin}, {Spergel}, {Staggs}, {Swetz}, {Switzer}, {Thornton}, {Trac},
  {Tucker}, {Warne}, {Wollack}, \& {Zhao}}]{dun11}
{Dunkley}, J., {Hlozek}, R., {Sievers}, J., {et~al.} 2011, \apj, 739, 52

\bibitem[{{Ebeling} {et~al.}(2000){Ebeling}, {Edge}, {Allen}, {Crawford},
  {Fabian}, \& {Huchra}}]{ebe00}
{Ebeling}, H., {Edge}, A.~C., {Allen}, S.~W., {et~al.} 2000, VizieR Online Data
  Catalog, 731, 80333

\bibitem[{{Ebeling} {et~al.}(2001){Ebeling}, {Edge}, \& {Henry}}]{ebe01}
{Ebeling}, H., {Edge}, A.~C., \& {Henry}, J.~P. 2001, \apj, 553, 668

\bibitem[{{Fischer} {et~al.}(1998){Fischer}, {Hasinger}, {Schwope}, {Brunner},
  {Boller}, {Tr{\"u}mper}, {Voges}, \& {Neizvestnyj}}]{fis98}
{Fischer}, J.-U., {Hasinger}, G., {Schwope}, A.~D., {et~al.} 1998,
  Astronomische Nachrichten, 319, 347

\bibitem[{{Fosalba} {et~al.}(2003){Fosalba}, {Gazta{\~n}aga}, \&
  {Castander}}]{fos03}
{Fosalba}, P., {Gazta{\~n}aga}, E., \& {Castander}, F.~J. 2003, \apjl, 597, L89

\bibitem[{{Gispert} {et~al.}(2000){Gispert}, {Lagache}, \& {Puget}}]{gis00}
{Gispert}, R., {Lagache}, G., \& {Puget}, J.~L. 2000, \aap, 360, 1

\bibitem[{{Gladders} \& {Yee}(2005)}]{gla05}
{Gladders}, M.~D. \& {Yee}, H.~K.~C. 2005, \apjs, 157, 1

\bibitem[{{G{\'o}rski} {et~al.}(2005){G{\'o}rski}, {Hivon}, {Banday},
  {Wandelt}, {Hansen}, {Reinecke}, \& {Bartelmann}}]{gor05}
{G{\'o}rski}, K.~M., {Hivon}, E., {Banday}, A.~J., {et~al.} 2005, \apj, 622,
  759

\bibitem[{{Hajian} {et~al.}(2013){Hajian}, {Battaglia}, {Spergel}, {Bond},
  {Pfrommer}, \& {Sievers}}]{haj13}
{Hajian}, A., {Battaglia}, N., {Spergel}, D.~N., {et~al.} 2013, ArXiv e-prints

\bibitem[{{Hasselfield} {et~al.}(2013){Hasselfield}, {Hilton}, {Marriage},
  {Addison}, {Barrientos}, {Battaglia}, {Battistelli}, {Bond}, {Crichton},
  {Das}, {Devlin}, {Dicker}, {Dunkley}, {D{\"u}nner}, {Fowler}, {Gralla},
  {Hajian}, {Halpern}, {Hincks}, {Hlozek}, {Hughes}, {Infante}, {Irwin},
  {Kosowsky}, {Marsden}, {Menanteau}, {Moodley}, {Niemack}, {Nolta}, {Page},
  {Partridge}, {Reese}, {Schmitt}, {Sehgal}, {Sherwin}, {Sievers}, {Sif{\'o}n},
  {Spergel}, {Staggs}, {Swetz}, {Switzer}, {Thornton}, {Trac}, \&
  {Wollack}}]{has13}
{Hasselfield}, M., {Hilton}, M., {Marriage}, T.~A., {et~al.} 2013, \jcap, 7, 8

\bibitem[{{Hern{\'a}ndez-Monteagudo} {et~al.}(2004){Hern{\'a}ndez-Monteagudo},
  {Genova-Santos}, \& {Atrio-Barandela}}]{her04}
{Hern{\'a}ndez-Monteagudo}, C., {Genova-Santos}, R., \& {Atrio-Barandela}, F.
  2004, \apjl, 613, L89

\bibitem[{{Hern{\'a}ndez-Monteagudo} {et~al.}(2006){Hern{\'a}ndez-Monteagudo},
  {Mac{\'{\i}}as-P{\'e}rez}, {Tristram}, \& {D{\'e}sert}}]{her06}
{Hern{\'a}ndez-Monteagudo}, C., {Mac{\'{\i}}as-P{\'e}rez}, J.~F., {Tristram},
  M., \& {D{\'e}sert}, F.-X. 2006, \aap, 449, 41

\bibitem[{{Hinshaw} {et~al.}(2007){Hinshaw}, {Nolta}, {Bennett}, {Bean},
  {Dor{\'e}}, {Greason}, {Halpern}, {Hill}, {Jarosik}, {Kogut}, {Komatsu},
  {Limon}, {Odegard}, {Meyer}, {Page}, {Peiris}, {Spergel}, {Tucker}, {Verde},
  {Weiland}, {Wollack}, \& {Wright}}]{hin07}
{Hinshaw}, G., {Nolta}, M.~R., {Bennett}, C.~L., {et~al.} 2007, \apjs, 170, 288

\bibitem[{{Hurier} {et~al.}(2014){Hurier}, {Aghanim}, \& {Douspis}}]{hur14a}
{Hurier}, G., {Aghanim}, N., \& {Douspis}, M. 2014, \aap, 568, A57

\bibitem[{{Hurier} {et~al.}(2013){Hurier}, {Mac{\'{\i}}as-P{\'e}rez}, \&
  {Hildebrandt}}]{hur13a}
{Hurier}, G., {Mac{\'{\i}}as-P{\'e}rez}, J.~F., \& {Hildebrandt}, S. 2013,
  \aap, 558, A118

\bibitem[{{Kalberla} {et~al.}(2005){Kalberla}, {Burton}, {Hartmann}, {Arnal},
  {Bajaja}, {Morras}, \& {P{\"o}ppel}}]{kal05}
{Kalberla}, P.~M.~W., {Burton}, W.~B., {Hartmann}, D., {et~al.} 2005, \aap,
  440, 775

\bibitem[{{Koester} {et~al.}(2007){Koester}, {McKay}, {Annis}, {Wechsler},
  {Evrard}, {Bleem}, {Becker}, {Johnston}, {Sheldon}, {Nichol}, {Miller},
  {Scranton}, {Bahcall}, {Barentine}, {Brewington}, {Brinkmann}, {Harvanek},
  {Kleinman}, {Krzesinski}, {Long}, {Nitta}, {Schneider}, {Sneddin}, {Voges},
  \& {York}}]{koe07}
{Koester}, B.~P., {McKay}, T.~A., {Annis}, J., {et~al.} 2007, \apj, 660, 239

\bibitem[{{Komatsu} \& {Kitayama}(1999)}]{kom99}
{Komatsu}, E. \& {Kitayama}, T. 1999, \apjl, 526, L1

\bibitem[{{Komatsu} \& {Seljak}(2002)}]{kom02}
{Komatsu}, E. \& {Seljak}, U. 2002, \mnras, 336, 1256

\bibitem[{{Komatsu} {et~al.}(2011){Komatsu}, {Smith}, {Dunkley}, {Bennett},
  {Gold}, {Hinshaw}, {Jarosik}, {Larson}, {Nolta}, {Page}, {Spergel},
  {Halpern}, {Hill}, {Kogut}, {Limon}, {Meyer}, {Odegard}, {Tucker}, {Weiland},
  {Wollack}, \& {Wright}}]{kom11}
{Komatsu}, E., {Smith}, K.~M., {Dunkley}, J., {et~al.} 2011, \apjs, 192, 18

\bibitem[{{Lieu} {et~al.}(2006){Lieu}, {Mittaz}, \& {Zhang}}]{lie06}
{Lieu}, R., {Mittaz}, J.~P.~D., \& {Zhang}, S.-N. 2006, \apj, 648, 176

\bibitem[{{Marriage} {et~al.}(2011){Marriage}, {Acquaviva}, {Ade}, {Aguirre},
  {Amiri}, {Appel}, {Barrientos}, {Battistelli}, {Bond}, {Brown}, {Burger},
  {Chervenak}, {Das}, {Devlin}, {Dicker}, {Bertrand Doriese}, {Dunkley},
  {D{\"u}nner}, {Essinger-Hileman}, {Fisher}, {Fowler}, {Hajian}, {Halpern},
  {Hasselfield}, {Hern{\'a}ndez-Monteagudo}, {Hilton}, {Hilton}, {Hincks},
  {Hlozek}, {Huffenberger}, {Handel Hughes}, {Hughes}, {Infante}, {Irwin},
  {Baptiste Juin}, {Kaul}, {Klein}, {Kosowsky}, {Lau}, {Limon}, {Lin},
  {Lupton}, {Marsden}, {Martocci}, {Mauskopf}, {Menanteau}, {Moodley},
  {Moseley}, {Netterfield}, {Niemack}, {Nolta}, {Page}, {Parker}, {Partridge},
  {Quintana}, {Reese}, {Reid}, {Sehgal}, {Sherwin}, {Sievers}, {Spergel},
  {Staggs}, {Swetz}, {Switzer}, {Thornton}, {Trac}, {Tucker}, {Warne},
  {Wilson}, {Wollack}, \& {Zhao}}]{mar11}
{Marriage}, T.~A., {Acquaviva}, V., {Ade}, P.~A.~R., {et~al.} 2011, \apj, 737,
  61

\bibitem[{{Mauch} {et~al.}(2003){Mauch}, {Murphy}, {Buttery}, {Curran},
  {Hunstead}, {Piestrzynski}, {Robertson}, \& {Sadler}}]{mau03}
{Mauch}, T., {Murphy}, T., {Buttery}, H.~J., {et~al.} 2003, \mnras, 342, 1117

\bibitem[{{Mauch} {et~al.}(2008){Mauch}, {Murphy}, {Buttery}, {Curran},
  {Hunstead}, {Piestrzynski}, {Ropbertson}, \& {Sadler}}]{mau08}
{Mauch}, T., {Murphy}, T., {Buttery}, H.~J., {et~al.} 2008, VizieR Online Data
  Catalog, 8081, 0

\bibitem[{{Melin} {et~al.}(2011){Melin}, {Bartlett}, {Delabrouille}, {Arnaud},
  {Piffaretti}, \& {Pratt}}]{mel11}
{Melin}, J.-B., {Bartlett}, J.~G., {Delabrouille}, J., {et~al.} 2011, \aap,
  525, A139

\bibitem[{{Mewe} {et~al.}(1985){Mewe}, {Gronenschild}, \& {van den
  Oord}}]{mew85}
{Mewe}, R., {Gronenschild}, E.~H.~B.~M., \& {van den Oord}, G.~H.~J. 1985,
  \aaps, 62, 197

\bibitem[{{Mo} \& {White}(1996)}]{mo96}
{Mo}, H.~J. \& {White}, S.~D.~M. 1996, \mnras, 282, 347

\bibitem[{{Morrison} \& {McCammon}(1983)}]{mor83}
{Morrison}, R. \& {McCammon}, D. 1983, \apj, 270, 119

\bibitem[{{Myers} {et~al.}(2004){Myers}, {Shanks}, {Outram}, {Frith}, \&
  {Wolfendale}}]{mye04}
{Myers}, A.~D., {Shanks}, T., {Outram}, P.~J., {Frith}, W.~J., \& {Wolfendale},
  A.~W. 2004, \mnras, 347, L67

\bibitem[{{Piffaretti} {et~al.}(2011){Piffaretti}, {Arnaud}, {Pratt},
  {Pointecouteau}, \& {Melin}}]{pif11}
{Piffaretti}, R., {Arnaud}, M., {Pratt}, G.~W., {Pointecouteau}, E., \&
  {Melin}, J.-B. 2011, \aap, 534, A109

\bibitem[{{Planck Collaboration early results. VIII}(2011)}]{planckESZ}
{Planck Collaboration early results. VIII}. 2011, \aap, 536, A8

\bibitem[{{Planck Collaboration early results. X}(2011)}]{planckxsz}
{Planck Collaboration early results. X}. 2011, \aap, 536, A10

\bibitem[{{Planck Collaboration early results. XI}(2011)}]{planckSL}
{Planck Collaboration early results. XI}. 2011, \aap, 536, A11

\bibitem[{{Planck Collaboration Int. results. VII}(2013)}]{planckRS}
{Planck Collaboration Int. results. VII}. 2013, \aap, 550, A133

\bibitem[{{Planck Collaboration results. I}(2014)}]{PlanckOVER}
{Planck Collaboration results. I}. 2014, \aap, 571, A1

\bibitem[{{Planck Collaboration results. IX}(2014)}]{planckBP}
{Planck Collaboration results. IX}. 2014, \aap, 571, A9

\bibitem[{{Planck Collaboration results. VII}(2014)}]{planckBEAMS}
{Planck Collaboration results. VII}. 2014, \aap, 571, A7

\bibitem[{{Planck Collaboration results. VIII}(2014)}]{PlanckCAL}
{Planck Collaboration results. VIII}. 2014, \aap, 571, A8

\bibitem[{{Planck Collaboration results. XII}(2014)}]{planckCOMP}
{Planck Collaboration results. XII}. 2014, \aap, 571, A12

\bibitem[{{Planck Collaboration results. XIII}(2013)}]{PlanckCO}
{Planck Collaboration results. XIII}. 2013, e-prints ArXiv: 1303.5073

\bibitem[{{Planck Collaboration results. XVI}(2014)}]{planckPAR}
{Planck Collaboration results. XVI}. 2014, \aap, 571, A16

\bibitem[{{Planck Collaboration results. XX}(2014)}]{PlanckSZC}
{Planck Collaboration results. XX}. 2014, \aap, 571, A20

\bibitem[{{Planck Collaboration results. XXI}(2014)}]{planckSZS}
{Planck Collaboration results. XXI}. 2014, \aap, 571, A21

\bibitem[{{Planck Collaboration results. XXIX}(2014)}]{PlanckPSZ}
{Planck Collaboration results. XXIX}. 2014, \aap, 571, A29

\bibitem[{{Planck Collaboration results. XXVIII}(2014)}]{planckPCC}
{Planck Collaboration results. XXVIII}. 2014, \aap, 571, A28

\bibitem[{{Pratt} {et~al.}(2009){Pratt}, {Croston}, {Arnaud}, \&
  {B{\"o}hringer}}]{pra09}
{Pratt}, G.~W., {Croston}, J.~H., {Arnaud}, M., \& {B{\"o}hringer}, H. 2009,
  \aap, 498, 361

\bibitem[{{Reichardt} {et~al.}(2012){Reichardt}, {Shaw}, {Zahn}, {Aird},
  {Benson}, {Bleem}, {Carlstrom}, {Chang}, {Cho}, {Crawford}, {Crites}, {de
  Haan}, {Dobbs}, {Dudley}, {George}, {Halverson}, {Holder}, {Holzapfel},
  {Hoover}, {Hou}, {Hrubes}, {Joy}, {Keisler}, {Knox}, {Lee}, {Leitch},
  {Lueker}, {Luong-Van}, {McMahon}, {Mehl}, {Meyer}, {Millea}, {Mohr},
  {Montroy}, {Natoli}, {Padin}, {Plagge}, {Pryke}, {Ruhl}, {Schaffer},
  {Shirokoff}, {Spieler}, {Staniszewski}, {Stark}, {Story}, {van Engelen},
  {Vanderlinde}, {Vieira}, \& {Williamson}}]{rei12}
{Reichardt}, C.~L., {Shaw}, L., {Zahn}, O., {et~al.} 2012, \apj, 755, 70

\bibitem[{{Reichert} {et~al.}(2011){Reichert}, {B{\"o}hringer}, {Fassbender},
  \& {M{\"u}hlegger}}]{rei11}
{Reichert}, A., {B{\"o}hringer}, H., {Fassbender}, R., \& {M{\"u}hlegger}, M.
  2011, \aap, 535, A4

\bibitem[{{Rykoff} {et~al.}(2013){Rykoff}, {Rozo}, {Busha}, {Cunha},
  {Finoguenov}, {Evrard}, {Hao}, {Koester}, {Leauthaud}, {Nord}, {Pierre},
  {Reddick}, {Sadibekova}, {Sheldon}, \& {Wechsler}}]{ryk13}
{Rykoff}, E.~S., {Rozo}, E., {Busha}, M.~T., {et~al.} 2013, ArXiv e-prints

\bibitem[{{Sehgal} {et~al.}(2011){Sehgal}, {Trac}, {Acquaviva}, {Ade},
  {Aguirre}, {Amiri}, {Appel}, {Barrientos}, {Battistelli}, {Bond}, {Brown},
  {Burger}, {Chervenak}, {Das}, {Devlin}, {Dicker}, {Bertrand Doriese},
  {Dunkley}, {D{\"u}nner}, {Essinger-Hileman}, {Fisher}, {Fowler}, {Hajian},
  {Halpern}, {Hasselfield}, {Hern{\'a}ndez-Monteagudo}, {Hilton}, {Hilton},
  {Hincks}, {Hlozek}, {Holtz}, {Huffenberger}, {Hughes}, {Hughes}, {Infante},
  {Irwin}, {Jones}, {Baptiste Juin}, {Klein}, {Kosowsky}, {Lau}, {Limon},
  {Lin}, {Lupton}, {Marriage}, {Marsden}, {Martocci}, {Mauskopf}, {Menanteau},
  {Moodley}, {Moseley}, {Netterfield}, {Niemack}, {Nolta}, {Page}, {Parker},
  {Partridge}, {Reid}, {Sherwin}, {Sievers}, {Spergel}, {Staggs}, {Swetz},
  {Switzer}, {Thornton}, {Tucker}, {Warne}, {Wollack}, \& {Zhao}}]{seh11}
{Sehgal}, N., {Trac}, H., {Acquaviva}, V., {et~al.} 2011, \apj, 732, 44

\bibitem[{{Shirokoff} {et~al.}(2011){Shirokoff}, {Reichardt}, {Shaw}, {Millea},
  {Ade}, {Aird}, {Benson}, {Bleem}, {Carlstrom}, {Chang}, {Cho}, {Crawford},
  {Crites}, {de Haan}, {Dobbs}, {Dudley}, {George}, {Halverson}, {Holder},
  {Holzapfel}, {Hrubes}, {Joy}, {Keisler}, {Knox}, {Lee}, {Leitch}, {Lueker},
  {Luong-Van}, {McMahon}, {Mehl}, {Meyer}, {Mohr}, {Montroy}, {Padin},
  {Plagge}, {Pryke}, {Ruhl}, {Schaffer}, {Spieler}, {Staniszewski}, {Stark},
  {Story}, {Vanderlinde}, {Vieira}, {Williamson}, \& {Zahn}}]{shi11}
{Shirokoff}, E., {Reichardt}, C.~L., {Shaw}, L., {et~al.} 2011, \apj, 736, 61

\bibitem[{{Sievers} {et~al.}(2013){Sievers}, {Hlozek}, {Nolta}, {Acquaviva},
  {Addison}, {Ade}, {Aguirre}, {Amiri}, {Appel}, {Barrientos}, {Battistelli},
  {Battaglia}, {Bond}, {Brown}, {Burger}, {Calabrese}, {Chervenak}, {Crichton},
  {Das}, {Devlin}, {Dicker}, {Bertrand Doriese}, {Dunkley}, {D{\"u}nner},
  {Essinger-Hileman}, {Faber}, {Fisher}, {Fowler}, {Gallardo}, {Gordon},
  {Gralla}, {Hajian}, {Halpern}, {Hasselfield}, {Hern{\'a}ndez-Monteagudo},
  {Hill}, {Hilton}, {Hilton}, {Hincks}, {Holtz}, {Huffenberger}, {Hughes},
  {Hughes}, {Infante}, {Irwin}, {Jacobson}, {Johnstone}, {Baptiste Juin},
  {Kaul}, {Klein}, {Kosowsky}, {Lau}, {Limon}, {Lin}, {Louis}, {Lupton},
  {Marriage}, {Marsden}, {Martocci}, {Mauskopf}, {McLaren}, {Menanteau},
  {Moodley}, {Moseley}, {Netterfield}, {Niemack}, {Page}, {Page}, {Parker},
  {Partridge}, {Plimpton}, {Quintana}, {Reese}, {Reid}, {Rojas}, {Sehgal},
  {Sherwin}, {Schmitt}, {Spergel}, {Staggs}, {Stryzak}, {Swetz}, {Switzer},
  {Thornton}, {Trac}, {Tucker}, {Uehara}, {Visnjic}, {Warne}, {Wilson},
  {Wollack}, {Zhao}, \& {Zunckel}}]{sie13}
{Sievers}, J.~L., {Hlozek}, R.~A., {Nolta}, M.~R., {et~al.} 2013, \jcap, 10, 60

\bibitem[{{Sunyaev} \& {Zeldovich}(1969)}]{sun69}
{Sunyaev}, R.~A. \& {Zeldovich}, Y.~B. 1969, \nat, 223, 721

\bibitem[{{Sunyaev} \& {Zeldovich}(1972)}]{sun72}
{Sunyaev}, R.~A. \& {Zeldovich}, Y.~B. 1972, Comments on Astrophysics and Space
  Physics, 4, 173

\bibitem[{{Taburet} {et~al.}(2011){Taburet}, {Hern{\'a}ndez-Monteagudo},
  {Aghanim}, {Douspis}, \& {Sunyaev}}]{tab11}
{Taburet}, N., {Hern{\'a}ndez-Monteagudo}, C., {Aghanim}, N., {Douspis}, M., \&
  {Sunyaev}, R.~A. 2011, \mnras, 418, 2207

\bibitem[{{The Dark Energy Survey Collaboration}(2005)}]{DES05}
{The Dark Energy Survey Collaboration}. 2005, ArXiv Astrophysics e-prints

\bibitem[{{Tinker} {et~al.}(2008){Tinker}, {Kravtsov}, {Klypin}, {Abazajian},
  {Warren}, {Yepes}, {Gottl{\"o}ber}, \& {Holz}}]{tin08}
{Tinker}, J., {Kravtsov}, A.~V., {Klypin}, A., {et~al.} 2008, \apj, 688, 709

\bibitem[{{Tristram} {et~al.}(2005){Tristram}, {Mac{\'{\i}}as-P{\'e}rez},
  {Renault}, \& {Santos}}]{tri05}
{Tristram}, M., {Mac{\'{\i}}as-P{\'e}rez}, J.~F., {Renault}, C., \& {Santos},
  D. 2005, \mnras, 358, 833

\bibitem[{{Vanderlinde} {et~al.}(2010){Vanderlinde}, {Crawford}, {de Haan},
  {Dudley}, {Shaw}, {Ade}, {Aird}, {Benson}, {Bleem}, {Brodwin}, {Carlstrom},
  {Chang}, {Crites}, {Desai}, {Dobbs}, {Foley}, {George}, {Gladders}, {Hall},
  {Halverson}, {High}, {Holder}, {Holzapfel}, {Hrubes}, {Joy}, {Keisler},
  {Knox}, {Lee}, {Leitch}, {Loehr}, {Lueker}, {Marrone}, {McMahon}, {Mehl},
  {Meyer}, {Mohr}, {Montroy}, {Ngeow}, {Padin}, {Plagge}, {Pryke}, {Reichardt},
  {Rest}, {Ruel}, {Ruhl}, {Schaffer}, {Shirokoff}, {Song}, {Spieler},
  {Stalder}, {Staniszewski}, {Stark}, {Stubbs}, {van Engelen}, {Vieira},
  {Williamson}, {Yang}, {Zahn}, \& {Zenteno}}]{van10}
{Vanderlinde}, K., {Crawford}, T.~M., {de Haan}, T., {et~al.} 2010, \apj, 722,
  1180

\bibitem[{{Vikhlinin} {et~al.}(2009){Vikhlinin}, {Kravtsov}, {Burenin},
  {Ebeling}, {Forman}, {Hornstrup}, {Jones}, {Murray}, {Nagai}, {Quintana}, \&
  {Voevodkin}}]{vik09}
{Vikhlinin}, A., {Kravtsov}, A.~V., {Burenin}, R.~A., {et~al.} 2009, \apj, 692,
  1060

\bibitem[{{Voges} {et~al.}(1999){Voges}, {Aschenbach}, {Boller},
  {Br{\"a}uninger}, {Briel}, {Burkert}, {Dennerl}, {Englhauser}, {Gruber},
  {Haberl}, {Hartner}, {Hasinger}, {K{\"u}rster}, {Pfeffermann}, {Pietsch},
  {Predehl}, {Rosso}, {Schmitt}, {Tr{\"u}mper}, \& {Zimmermann}}]{vog99}
{Voges}, W., {Aschenbach}, B., {Boller}, T., {et~al.} 1999, \aap, 349, 389

\end{thebibliography}

\end{document}